\documentclass[conference,a4paper]{IEEEtran}

\usepackage{mdwmath}
\usepackage{mdwtab}
\usepackage{graphicx}
\usepackage{amsmath}
\usepackage{latexsym}
\usepackage{amssymb}
\usepackage{comment}
\usepackage{cite}
\usepackage{url}
\usepackage[T3,T1]{fontenc}
\DeclareSymbolFont{tipa}{T3}{cmr}{m}{n}
\DeclareMathAccent{\invbreve}{\mathalpha}{tipa}{16}

\global\long\def\L{\mathsf{L}}
\global\long\def\D{\mathsf{D}}
\global\long\def\Sgen{\mathcal{S}_\mathsf{gen}}
\global\long\def\Kgen{\mathcal{K}_\mathsf{gen}}
\global\long\def\A{\mathcal{A}}

\newcommand{\calVarX}{{\cal X}}

\newcommand{\tX}{\overline{X}}

\newcommand{\E}{\mbox{\bf E}}

\newcommand{\Dist}[1]{{#1}}

\newcommand{\prmtA}{\mu}
\newcommand{\prmtB}{\bar{\mu}}

\begin{document}
%
\title{
Information Theoretic Security for 
Side-Channel Attacks to the Shannon Cipher System
}
\author{%
	\IEEEauthorblockN{Yasutada Oohama and Bagus Santoso}
	\IEEEauthorblockA{University of Electro-Communications, Tokyo, Japan\\ 
	Email: \url{{oohama,santoso.bagus}@uec.ac.jp}}
%
}
\maketitle

\begin{abstract}
We study side-channel attacks for the Shannon cipher system. 
To pose side channel-attacks to the Shannon cipher system, 
we regard them as a signal estimation via encoded data from 
two distributed sensors. This can be formulated as the one helper 
source coding problem posed and investigated
by Ahlswede, K\"orner(1975), and Wyner(1975). We further investigate 
the posed problem to derive new secrecy bounds. Our results 
are derived by a coupling of the result Watanabe and Oohama(2012) 
obtained on bounded storage eavesdropper with the exponential strong converse 
theorem Oohama(2015) established for the one helper 
source coding problem.
\end{abstract}
%
\IEEEpeerreviewmaketitle

\newcommand{\qed}{\hfill$\square$}
\newcommand{\suchthat}{\mbox{~s.t.~}}
\newcommand{\markov}{\leftrightarrow}

\newcommand{\argmax}{\mathop{\rm argmax}\limits}
\newcommand{\argmin}{\mathop{\rm argmin}\limits}

\newcommand{\ExP}{\rm e}

\newcommand{\Cls}{class NL}
\newcommand{\vSpa}{\vspace{0.3mm}}
\newcommand{\Prmt}{\zeta}
\newcommand{\pj}{\omega_n}

\newfont{\bg}{cmr10 scaled \magstep4}
\newcommand{\bigzerol}{\smash{\hbox{\bg 0}}}
\newcommand{\bigzerou}{\smash{\lower1.7ex\hbox{\bg 0}}}
\newcommand{\nbn}{\frac{1}{n}}
\newcommand{\ra}{\rightarrow}
\newcommand{\la}{\leftarrow}
\newcommand{\ldo}{\ldots}
\newcommand{\typi}{A_{\epsilon }^{n}}
\newcommand{\bx}{\hspace*{\fill}$\Box$}
\newcommand{\pa}{\vert}
\newcommand{\ignore}[1]{}


\newtheorem{proposition}{Proposition}
\newtheorem{definition}{Definition}
\newtheorem{theorem}{Theorem}
\newtheorem{lemma}{Lemma}
\newtheorem{corollary}{Corollary}
\newtheorem{remark}{Remark}
\newtheorem{property}{Property}

\newcommand{\defeq}{:=}

\newcommand{\Qed}{\hbox{\rule[-2pt]{3pt}{6pt}}}
\newcommand{\beq}{\begin{equation}}
\newcommand{\eeq}{\end{equation}}
\newcommand{\beqa}{\begin{eqnarray}}
\newcommand{\eeqa}{\end{eqnarray}}
\newcommand{\beqno}{\begin{eqnarray*}}
\newcommand{\eeqno}{\end{eqnarray*}}
\newcommand{\ba}{\begin{array}}
\newcommand{\ea}{\end{array}}

\newcommand{\vc}[1]{\mbox{\boldmath $#1$}}
\newcommand{\lvc}[1]{\mbox{\scriptsize \boldmath $#1$}}
\newcommand{\svc}[1]{\mbox{\scriptsize\boldmath $#1$}}

\newcommand{\wh}{\widehat}
\newcommand{\wt}{\widetilde}
\newcommand{\ts}{\textstyle}
\newcommand{\ds}{\displaystyle}
\newcommand{\scs}{\scriptstyle}
\newcommand{\vep}{\varepsilon}
\newcommand{\rhp}{\rightharpoonup}
\newcommand{\cl}{\circ\!\!\!\!\!-}
\newcommand{\bcs}{\dot{\,}.\dot{\,}}
\newcommand{\eqv}{\Leftrightarrow}
\newcommand{\leqv}{\Longleftrightarrow}

\newcommand{\irr}[1]{{\color[named]{Red}#1\normalcolor}}

\newcommand{\hugel}{{\arraycolsep 0mm
                    \left\{\ba{l}{\,}\\{\,}\ea\right.\!\!}}
\newcommand{\Hugel}{{\arraycolsep 0mm
                    \left\{\ba{l}{\,}\\{\,}\\{\,}\ea\right.\!\!}}
\newcommand{\HUgel}{{\arraycolsep 0mm
                    \left\{\ba{l}{\,}\\{\,}\\{\,}\vspace{-1mm}
                    \\{\,}\ea\right.\!\!}}
\newcommand{\huger}{{\arraycolsep 0mm
                    \left.\ba{l}{\,}\\{\,}\ea\!\!\right\}}}
\newcommand{\Huger}{{\arraycolsep 0mm
                    \left.\ba{l}{\,}\\{\,}\\{\,}\ea\!\!\right\}}}
\newcommand{\HUger}{{\arraycolsep 0mm
                    \left.\ba{l}{\,}\\{\,}\\{\,}\vspace{-1mm}
                    \\{\,}\ea\!\!\right\}}}

\newcommand{\hugebl}{{\arraycolsep 0mm
                    \left[\ba{l}{\,}\\{\,}\ea\right.\!\!}}
\newcommand{\Hugebl}{{\arraycolsep 0mm
                    \left[\ba{l}{\,}\\{\,}\\{\,}\ea\right.\!\!}}
\newcommand{\HUgebl}{{\arraycolsep 0mm
                    \left[\ba{l}{\,}\\{\,}\\{\,}\vspace{-1mm}
                    \\{\,}\ea\right.\!\!}}
\newcommand{\hugebr}{{\arraycolsep 0mm
                    \left.\ba{l}{\,}\\{\,}\ea\!\!\right]}}
\newcommand{\Hugebr}{{\arraycolsep 0mm
                    \left.\ba{l}{\,}\\{\,}\\{\,}\ea\!\!\right]}}
\newcommand{\HUgebr}{{\arraycolsep 0mm
                    \left.\ba{l}{\,}\\{\,}\\{\,}\vspace{-1mm}
                    \\{\,}\ea\!\!\right]}}

\newcommand{\hugecl}{{\arraycolsep 0mm
                    \left(\ba{l}{\,}\\{\,}\ea\right.\!\!}}
\newcommand{\Hugecl}{{\arraycolsep 0mm
                    \left(\ba{l}{\,}\\{\,}\\{\,}\ea\right.\!\!}}
\newcommand{\hugecr}{{\arraycolsep 0mm
                    \left.\ba{l}{\,}\\{\,}\ea\!\!\right)}}
\newcommand{\Hugecr}{{\arraycolsep 0mm
                    \left.\ba{l}{\,}\\{\,}\\{\,}\ea\!\!\right)}}

\newcommand{\hugepl}{{\arraycolsep 0mm
                    \left|\ba{l}{\,}\\{\,}\ea\right.\!\!}}
\newcommand{\Hugepl}{{\arraycolsep 0mm
                    \left|\ba{l}{\,}\\{\,}\\{\,}\ea\right.\!\!}}
\newcommand{\hugepr}{{\arraycolsep 0mm
                    \left.\ba{l}{\,}\\{\,}\ea\!\!\right|}}
\newcommand{\Hugepr}{{\arraycolsep 0mm
                    \left.\ba{l}{\,}\\{\,}\\{\,}\ea\!\!\right|}}

\newcommand{\MEq}[1]{\stackrel{
{\rm (#1)}}{=}}

\newcommand{\MLeq}[1]{\stackrel{
{\rm (#1)}}{\leq}}

\newcommand{\ML}[1]{\stackrel{
{\rm (#1)}}{<}}

\newcommand{\MGeq}[1]{\stackrel{
{\rm (#1)}}{\geq}}

\newcommand{\MG}[1]{\stackrel{
{\rm (#1)}}{>}}

\newcommand{\MPreq}[1]{\stackrel{
{\rm (#1)}}{\preceq}}

\newcommand{\MSueq}[1]{\stackrel{
{\rm (#1)}}{\succeq}}

\newcommand{\MSubeq}[1]{\stackrel{
{\rm (#1)}}{\subseteq}}

\newcommand{\MSupeq}[1]{\stackrel{
{\rm (#1)}}{\supseteq}}

\newcommand{\MRarrow}[1]{\stackrel{
{\rm (#1)}}{\Rightarrow}}

\newcommand{\MLarrow}[1]{\stackrel{
{\rm (#1)}}{\Leftarrow}}

\newcommand{\SZZpp}{
}

\newcommand{\vcc}{{c}^n}
\newcommand{\vck}{{k}^n}
\newcommand{\vcx}{{x}^n}
\newcommand{\vcy}{{y}^n}
\newcommand{\vcz}{{z}^n}
\newcommand{\vckone}{{k}_1^n}
\newcommand{\vcktwo}{{k}_2^n}
\newcommand{\vcxone}{{x}^n}

\newcommand{\vcxtwo}{{x}_2^n}
\newcommand{\vcyone}{{y}_1^n}
\newcommand{\vcytwo}{{y}_2^n}

\newcommand{\cvcx}{\check{x}^n}
\newcommand{\cvcy}{\check{y}^n}
\newcommand{\cvcz}{\check{z}^n}
\newcommand{\cvcxone}{\check{x}^n}

\newcommand{\cvcxtwo}{\check{x}_2^n}

\newcommand{\hvcx}{\widehat{x}^n}
\newcommand{\hvcy}{\widehat{y}^n}
\newcommand{\hvcz}{\widehat{z}^n}
\newcommand{\hvckone}{\widehat{k}_1^n}
\newcommand{\hvcktwo}{\widehat{k}_2^n}

\newcommand{\hvcxone}{\widehat{x}^n}

\newcommand{\hvcxtwo}{\widehat{x}_2^n}

\newcommand{\lvcc}{{c}^n}
\newcommand{\lvck}{{k}^n}
\newcommand{\lvcx}{{x}^n}
\newcommand{\lvcy}{{y}^n}
\newcommand{\lvcz}{{z}^n}

\newcommand{\lvckone}{{k}_1^n}
\newcommand{\lvcktwo}{{k}_2^n}
\newcommand{\lvcxone}{{x}^n}

\newcommand{\lvcxtwo}{{x}_2^n}
\newcommand{\lvcyone}{{y}_1^n}
\newcommand{\lvcytwo}{{y}_2^n}

\newcommand{\clvcxone}{\check{x}^n}

\newcommand{\clvcxtwo}{\check{x}_2^n}

\newcommand{\hlvckone}{\widehat{k}_1^n}
\newcommand{\hlvcktwo}{\widehat{k}_2^n}

\newcommand{\hlvcxone}{\widehat{x}^n}

\newcommand{\hlvcxtwo}{\widehat{x}_2^n}

\newcommand{\rvcc}{{C}^n}
\newcommand{\rvck}{{K}^n}
\newcommand{\rvcx}{{X}^n}
\newcommand{\rvcy}{{Y}^n}
\newcommand{\rvcz}{{Z}^n}
\newcommand{\rvccone}{{C}_1^n}
\newcommand{\rvcctwo}{{C}_2^n}
\newcommand{\rvckone}{{K}_1^n}
\newcommand{\rvcktwo}{{K}_2^n}
\newcommand{\rvcxone}{{X}^n}

\newcommand{\rvcxtwo}{{X}_2^n}
\newcommand{\rvcyone}{{Y}_1^n}
\newcommand{\rvcytwo}{{Y}_2^n}
\newcommand{\hrvcx}{\widehat{X}^n}
\newcommand{\hrvcxone}{\widehat{X}_1^n}
\newcommand{\hrvcxtwo}{\widehat{X}_2^n}

\newcommand{\lrvcc}{{C}^n}
\newcommand{\lrvck}{{K}^n}
\newcommand{\lrvcx}{{X}^n}
\newcommand{\lrvcy}{{Y}^n}
\newcommand{\lrvcz}{{Z}^n}
\newcommand{\lrvckone}{{K}_1^n}
\newcommand{\lrvcktwo}{{K}_2^n}

\newcommand{\lrvcxone}{{X}^n}
\newcommand{\lrvcxtwo}{{X}_2^n}
\newcommand{\lrvcyone}{{Y}_1^n}
\newcommand{\lrvcytwo}{{Y}_2^n}
\newcommand{\rvcci}{{C}_i^n}
\newcommand{\rvcki}{{K}_i^n}
\newcommand{\rvcxi}{{X}_i^n}
\newcommand{\rvcyi}{{Y}_i^n}
\newcommand{\hrvcxi}{\widehat{X}_i^n}
\newcommand{\vcki}{{k}_i^n}
\newcommand{\vcsi}{{s}_i^n}
\newcommand{\vcti}{{t}_i^n}
\newcommand{\vcvi}{{v}_i^n}
\newcommand{\vcwi}{{w}_i^n}
\newcommand{\vcxi}{{x}_i^n}
\newcommand{\vcyi}{{y}_i^n}

\newcommand{\vcs}{{s}^n}
\newcommand{\vct}{{t}^n}
\newcommand{\vcv}{{v}^n}
\newcommand{\vcw}{{w}^n}
%
%

\newcommand{\SZZ}{

\newcommand{\vcc}{{\vc c}}
\newcommand{\vck}{{\vc k}}
\newcommand{\vcx}{{\vc x}}
\newcommand{\vcy}{{\vc y}}
\newcommand{\vcz}{{\vc z}}
\newcommand{\vckone}{{\vc k}_1}
\newcommand{\vcktwo}{{\vc k}_2}
\newcommand{\vcxone}{{\vc x}_1}
\newcommand{\vcxtwo}{{\vc x}_2}
\newcommand{\vcyone}{{\vc y}_1}
\newcommand{\vcytwo}{{\vc y}_2}

\newcommand{\cvcx}{\check{\vc x}}
\newcommand{\cvcy}{\check{\vc y}}
\newcommand{\cvcz}{\check{\vc z}}
\newcommand{\cvcxone}{\check{\vc x}_1}
\newcommand{\cvcxtwo}{\check{\vc x}_2}

\newcommand{\hvcx}{\widehat{\vc x}}
\newcommand{\hvcy}{\widehat{\vc y}}
\newcommand{\hvcz}{\widehat{\vc z}}
\newcommand{\hvckone}{\widehat{\vc k}_1}
\newcommand{\hvcktwo}{\widehat{\vc k}_2}
\newcommand{\hvcxone}{\widehat{\vc x}_1}
\newcommand{\hvcxtwo}{\widehat{\vc x}_2}

\newcommand{\lvcc}{{c}}
\newcommand{\lvck}{{k}}
\newcommand{\lvcx}{{x}}
\newcommand{\lvcy}{{y}}
\newcommand{\lvcz}{{z}}

\newcommand{\lvckone}{{k}_1}
\newcommand{\lvcktwo}{{k}_2}

\newcommand{\lvcxone}{{x}}

\newcommand{\lvcxtwo}{{x}_2}
\newcommand{\lvcyone}{{y}_1}
\newcommand{\lvcytwo}{{y}_2}

\newcommand{\clvcxone}{\check{x}_1}
\newcommand{\clvcxtwo}{\check{x}_2}

\newcommand{\hlvckone}{\widehat{k}_1}
\newcommand{\hlvcktwo}{\widehat{k}_2}
\newcommand{\hlvcxone}{\widehat{x}_1}
\newcommand{\hlvcxtwo}{\widehat{x}_2}

\newcommand{\rvcc}{{\vc C}}
\newcommand{\rvck}{{\vc K}}
\newcommand{\rvcx}{{\vc X}}
\newcommand{\rvcy}{{\vc Y}}
\newcommand{\rvcz}{{\vc Z}}
\newcommand{\rvccone}{{\vc C}_1}
\newcommand{\rvcctwo}{{\vc C}_2}
\newcommand{\rvckone}{{\vc K}_1}
\newcommand{\rvcktwo}{{\vc K}_2}

\newcommand{\rvcxone}{{\vc X}}

\newcommand{\rvcxtwo}{{\vc X}_2}
\newcommand{\rvcyone}{{\vc Y}_1}
\newcommand{\rvcytwo}{{\vc Y}_2}
\newcommand{\hrvcxone}{\widehat{\vc X}_1}
\newcommand{\hrvcxtwo}{\widehat{\vc X}_2}

\newcommand{\lrvcc}{{C}}
\newcommand{\lrvck}{{K}}
\newcommand{\lrvcx}{{X}}
\newcommand{\lrvcy}{{Y}}
\newcommand{\lrvcz}{{Z}}
\newcommand{\lrvckone}{{K}_1}
\newcommand{\lrvcktwo}{{K}_2}
\newcommand{\lrvcxone}{{X}_1}
\newcommand{\lrvcxtwo}{{X}_2}
\newcommand{\lrvcyone}{{Y}_1}
\newcommand{\lrvcytwo}{{Y}_2}
\newcommand{\rvcci}{{\vc C}_i}
\newcommand{\rvcki}{{\vc K}_i}
\newcommand{\rvcxi}{{\vc X}_i}
\newcommand{\rvcyi}{{\vc Y}_i}
\newcommand{\hrvcxi}{\widehat{\vc X}_i}
\newcommand{\vcki}{{\vc k}_i}
\newcommand{\vcsi}{{\vc s}_i}
\newcommand{\vcti}{{\vc t}_i}
\newcommand{\vcvi}{{\vc v}_i}
\newcommand{\vcwi}{{\vc w}_i}
\newcommand{\vcxi}{{\vc x}_i}
\newcommand{\vcyi}{{\vc y}_i}

\newcommand{\vcs}{{\vc s}}
\newcommand{\vct}{{\vc t}}
\newcommand{\vcv}{{\vc v}}
\newcommand{\vcw}{{\vc w}}
}

\section{Introduction \label{sec:introduction}}

In this paper, we consider the problem of strengthening the security of  
communication in the Shannon cipher system when we have side channel 
attacks to the cryptosystem. 
Especially, we are interested on practical solutions with minimum
modifications which can be applied even on  already running systems.

More precisely, we consider a cryptosystem described
as follows: a source $X$ is encrypted in a node to $C$ 
using secret key $K$. The cipher text $C$ is sent 
through a public communication channel to a sink node, 
where $X$ is decrypted from $C$ using $K$. 
We suppose that an already running system 
has a potential secrecy/privacy
problem such that $X$ might be leaked to  
an adversary which is eavesdropping the public 
communication channel and is also using 
a side-channel providing some side information on 
$K$. 

To pose side channel-attacks to the Shannon cipher system, we regard 
them as a signal estimation via encoded data from two distributed 
sensors. This can be formulated as the one helper source coding problem 
posed and investigated by Ahlswede, K\"orner \cite{ahlswede:75} and Wyner 
\cite{wyner:75c}. 

We further investigate the posed problem to derive new secrecy bounds. 
Our results are derived by two previous results. One is the coding 
theorem Watanebe and Oohama \cite{watanabe2012privacy} obtained 
for the privacy amplification problem for bounded storage eavesdropper 
posed by them. The other is the exponential strong converse theorem  
Oohama \cite{oohama2015exponent} established for the one 
helper source coding problem.

\section{Problem Formulation}

\subsection{Preliminaries}

In this subsection, we show the basic notations and related consensus 
used in this paper. 

\noindent{}
\textit{Random Source of Information and Key: \ }
Let $X$ be a random variable from a finite set 
$\mathcal{X}$. 
Let $\{X_t\}_{t=1}^\infty$ be a stationary discrete
memoryless source(DMS) such that for each $t=1,2,\ldots$, 
$X_{t}$ takes values in finite set $\mathcal{X}$ 
and obeys the same distribution 
as that of $X$ denoted by 
${p}_{X}=\{{p}_{X} (x)\}_{x \in \mathcal{X}}$.
The stationary DMS $\{X_t\}_{t=1}^\infty$ 
is specified with ${p}_{X}$.
Also, let $K$ be a random variable taken from 
the same finite set $\mathcal{X}$ representing the key 
used for encryption.
Similarly, let $\{K_{t}\}_{t=1}^\infty$ be a stationary discrete
memoryless source such that for each $t=1,2,\ldots$, $K_{t}$ 
takes values in the finite set
$\mathcal{X}$ and obeys the same distribution as that of $K$ denoted by 
${p}_{K}=\{{p}_{K} (k)\}_{k\in\mathcal{X}}$.
The stationary DMS $\{K_t\}_{t=1}^\infty$ 
is specified with ${p}_{K}$. In this paper we assume that 
$p_K$ is the uniform distribution over ${\cal X}$.  

\noindent{}\textit{Random Variables and Sequences: \ }
We write the sequence of random variables with length $n$ 
from the information source as follows:
${\rvcx}\defeq X_{1}X_{2}\cdots X_{n}$. 
Similarly, the strings with length 
$n$ of $\mathcal{X}^n$ are written as 
${\vcx}\defeq x_{1}x_{2}\cdots
x_{n}\in\mathcal{X}^n$. 
For ${\vcx}\in \mathcal{X}^n$, 
${p}_{{\lrvcx}}({\vcx})$ stands for the 
probability of the occurrence of 
${\vcx}$. 
When the information source is memoryless 
specified with ${p}_{X}$, we have 
the following equation holds:
$$
{p}_{{\lrvcx}}({\vcx})=\prod_{t=1}^n {p}_{X}(x_t).
$$
In this case we write ${p}_{{\lrvcx}}({\vcx})$
as ${p}_{X}^n({\vcx})$. Similar notations are 
used for other random variables and sequences.

\noindent{}\emph{Consensus and Notations: }
Without loss of generality, throughout this paper,
we assume that $\mathcal{X}$ is a finite field.
The notation $\oplus$ is used to denote the field 
addition operation, while the notation $\ominus$ 
is used to denote the field subtraction operation, i.e., 
$a\ominus b = a \oplus (-b)$ for any elements 
$a,b \in {\cal X}$. Throughout 
this paper all logarithms are 
taken to the base natural.

\subsection{Basic System Description}

\begin{figure}[t]
\centering
\includegraphics[width=0.47\textwidth]{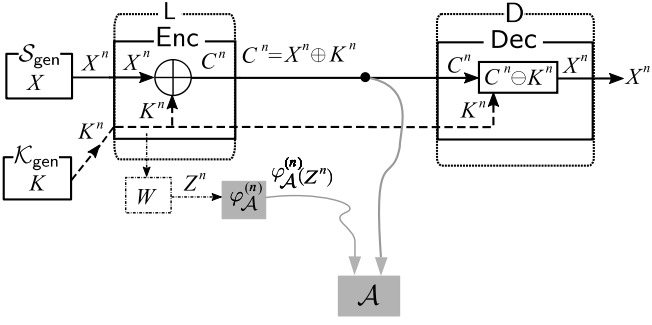}
\caption{Side-channel attacks to the Shannon cipher system.
\label{fig:main}}
\end{figure}

In this subsection we explain the basic system setting and 
basic adversarial model we consider in this paper.
First, let the information source and the key be generated 
independently by different parties
$\Sgen$ and $\Kgen$ respectively.
In our setting, we assume the followings.
\begin{itemize}
\item The random key ${\rvck}$ is generated by $\Kgen$
from uniform distribution.
\item The source is generated by $\Sgen$ and 
independent of the key.
\end{itemize}
Next, let the random source ${\rvcx}$ from $\Sgen$
be sent to the node $\mathsf{L}$.
And let the random key ${\rvck}$ from $\Kgen$
be also sent to $\mathsf{L}$.
Further settings of our system are described as follows. 
Those are also shown in Fig. \ref{fig:main}.
\begin{enumerate}
	\item \emph{Source Processing:} \ At the node 
	$\L$, ${\rvcx}$ is encrypted with 
        the key ${\rvck}$ using the encryption function $\mathsf{Enc}$.
        The ciphertext ${\rvcc}$ of ${\rvcx}$ is given by
        $$
        {\rvcc} \defeq \mathsf{Enc}({\rvcx})={\rvcx}\oplus {\rvck}.
        $$
	\item \emph{Transmission:} \ Next, the 
	ciphertext ${\rvcc}$ is sent to the 
        information processing center $\D$ through a \emph{public} 
        communication channel. Meanwhile, the key ${\rvck}$ 
        is sent to $\D$
	through a \emph{private} communication channel.

	\item \emph{Sink Node Processing:} \ In $\D$, we decrypt 
        the ciphertext ${\rvcc}$ using 
        the key ${\rvck}$ through the corresponding decryption 
        procedure $\mathsf{Dec}$ defined by 
	$\mathsf{Dec}({\rvcc})={\rvcc} \ominus {\rvck}$. 
        It is obvious that we can correctly reproduce the source 
        output $\rvcx$ from $\rvcc$ and $\rvck$ 
        by the decryption function $\mathsf{Dec}$.
\end{enumerate}

\noindent
\underline{\it Side-Channel Attacks by Eavesdropper Adversary:} 
An \emph{(eavesdropper) adversary} $\A$ eavesdrops the public 
communication channel in the system. The adversary $\A$ 
also uses a side information obtained by side-channel attacks. 
In this paper we introduce a {\it new theoretical model 
of side-channel attacks}, which is described as follows. 
Let ${\cal Z}$ be a finite set and let $W:{\cal X}\to {\cal Z}$ 
be a noisy channel. 
Let $Z$ be a channel output from $W$ for the input random variable $K$. 
We consider the discrete memoryless channel specified 
with $W$. Let $\rvcz \in {\cal Z}^n$ be a random variable obtained as 
the channel output by connecting $\rvck \in {\cal X}^n$ to 
the input of channel. We write a conditional 
distribution on $\rvcz$ given $\rvck$ as 
$$
W^n=
\left\{W^n(\vcz|\vck)\right
\}_{(\lvck,\lvcz)\in {\cal K}^n \times {\cal Z}^n}.
$$
Since the channel is memoryless, we have 
\beq
W^n({\vcz}|{\vck})=\prod_{t=1}^nW (z_t|k_t).
\label{eqn:sde0}
\eeq
On the above output $\rvcz$ of $W^n$ for the input $\rvck$, 
we assume the followings.
\begin{itemize}
\item The three random variables $X$, $K$ and $Z$, satisfy 
$X \perp (K,Z)$, which implies that $X^n \perp (K^n,Z^n)$.
\item $W$ is given 
in the system and the adversary ${\cal A}$ can not control $W$. 
\item By side-channel attacks, the adversary ${\cal A}$ 
can access $Z^n$. 
\end{itemize}
We next formulate side information the adversary ${\cal A}$ 
obtains by side-channel attacks. For each $n=1,2,\cdots$, 
let $\varphi_{\cal A}^{(n)}:{\cal Z}^n 
\to {\cal M}_{\cal A}^{(n)}$ be an encoder function. 
Set 
$
\varphi_{\cal A} \defeq \{\varphi_{\cal A}^{(n)}\}_{n=1,2,\cdots}.
$ 
Let 
$$ 
R_{\cal A}^{(n)}\defeq 
\frac{1}{n} \log ||\varphi_{\cal A}||
=\frac{1}{n} \log |{\cal M}_{\cal A}^{(n)}|
$$
be a rate of the encoder function $\varphi_{\cal A}^{(n)}$. 
For $R_{\cal A}>0$, we set 
$$
{\cal F}_{\cal A}^{(n)}(R_{\cal A})
\defeq \{ \varphi_{\cal A}^{(n)}: R_{\cal A}^{(n)} \leq R_{\cal A}\}.
$$ 
On encoded side information the adversary ${\cal A}$ obtains 
we assume the following.
\begin{itemize}
\item The adversary ${\cal A}$, having accessed $Z^n$, obtains 
the encoded additional information $\varphi_{\cal A}^{(n)}(\rvcz)$.
For each $n=1,2,\cdots$, the adversary ${\cal A}$ 
can design $\varphi_{\cal A}^{(n)}$. 
\item The sequence $\{R_{\cal A}^{(n)}\}_{n=1}^{\infty}$
must be upper bounded by a prescribed value. 
In other words, the adversary ${\cal A}$ must use $ \varphi_{\cal A}^{(n)}$ 
such that for some $R_{\cal A}$ and for any sufficiently large $n$, 
$\varphi_{\cal A}^{(n)}\in {\cal F}_{\cal A}^{(n)}(R_{\cal A})$. 
\end{itemize}
\noindent{}\textit{Validity of Our Theoretical Model:\ }
When the $|{\cal Z}|$ is not so large the adversary ${\cal A}$ may 
directly access to $Z^n$. On the contrary, as a real situation of side 
channel attacks we have often the case where the noisy version $Z^n$ of 
$K^n$ can be regarded as almost an analog random signal. In this case, 
$|{\cal Z}|$ is sufficiently large and the adversary ${\cal A}$ can not 
obtain $Z^n$ in a lossless form. Our theoretical model can address such 
situations of side channel attacks.
\begin{figure}[t] 
	\centering 
	\includegraphics[width=0.48\textwidth]
	{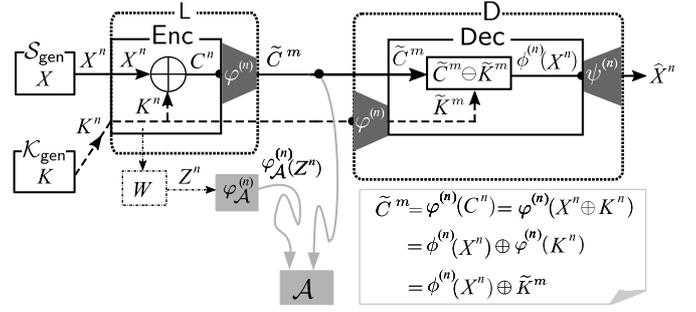}
	\caption{Our proposed solution: linear encoders 
	as privacy amplifiers.
\label{fig:solution}}%
\end{figure}

\subsection{Proposed Idea: Affine Encoder as Privacy Amplifier}
%
For each $n=1,2,\cdots$, let $\phi^{(n)}: {\cal X}^n \to {\cal X}^{m}$ 
be a linear mapping. We define the mapping $\phi^{(n)}$ by 
\beq
\phi^{(n)}({\vcx})
={\vcx} A \mbox{ for }{\vcx} \in {\cal X}^n,
\label{eq:homomorphica}
\eeq
where $A$ is a matrix with $n$ rows 
and $m$ columns. Entries of $A$ are 
from ${\cal X}$. We fix $b^{m}\in \mathcal{X}^{m}$. 
Define the mapping 
$\varphi^{(n)}: {\cal X}^n \to {\cal X}^{m}$
by 
\begin{align}
\varphi^{(n)}({\vck}):=&
\phi^{(n)}({\vck})\oplus b^{m}
\notag \\
=&{\vck} A \oplus b^{m}, 
\mbox{ for }{\vck} \in \mathcal{X}^n.
\label{eq:homomorphic}
\end{align}
The mapping $\varphi^{(n)}$ is called the affine 
mapping induced by the linear mapping $\phi^{(n)}$ and constant 
vector $b^{m}$ $\in{\cal X}^{m}$.
By the definition (\ref{eq:homomorphic}) of 
$\varphi^{(n)}$, those satisfy the following 
affine structure: 
\begin{align}
&\varphi^{(n)}({\vcy} \oplus {\vck})
({\vcx}\oplus {\vck})A\oplus b^{m}=
{\vcx}A\oplus({\vck}A\oplus b^{m})
\notag\\
&=\phi^{(n)}({\vcx})\oplus \varphi^{(n)}({\vck}),
\mbox{ for } {\vcx}, {\vck} \in {\cal X}^n.
\label{eq:affine}
\end{align}
Next, let $\psi^{(n)}$ be the corresponding decoder for 
$\phi^{(n)}$ such that
$
\psi^{(n)}: \mathcal{X}^{m} 
\rightarrow \mathcal{X}^{n}.
$
Note that $\psi^{(n)}$ does not have a linear structure in general. 

\noindent
\underline{\it Description of Proposed Procedure:} 
We describe the procedure of our privacy 
amplified system as follows. 
\begin{enumerate}
	\item\emph{Encoding of Ciphertext:} \ First, we use 
	$\varphi^{(n)}$ to encode the ciphertext $C^{n}=X^{n}\oplus K^{n}$
	Let $\widetilde{C}^{m}=\varphi^{(n)}({\rvcc})$.
	Then, instead of sending ${\rvcc}$, we send $\tilde{C}^{m}$ 
        to the public communication channel. By the affine 
        structure (\ref{eq:affine}) of encoder we have that
        \begin{align}
        &\widetilde{C}^{m}=\varphi^{(n)}({\rvcx}\oplus {\rvck})
        \notag \\
        &=\phi^{(n)}({\rvcx}) \oplus \varphi^{(n)}({\rvck})
        =\widetilde{X}^{m} \oplus \widetilde{K}^{m},
        \label{eqn:aSdzx} 
        \end{align}
        where we set 
        $
	\widetilde{X}^{m} \defeq \phi^{(n)}({\rvcx}),
        \widetilde{K}^{m} \defeq \varphi^{(n)}({\rvck}).
        $
	\item\emph{Decoding at Sink Node $\D$: \ }
	First, using the linear encoder
	$\varphi^{(n)}$,
	$\D$ encodes the key $\rvck$ received
	through private channel into
	$\widetilde{K}^{m}=$$(\varphi^{(n)}({\rvck})$.
	Receiving $\widetilde{C}^{m}$ from
	public communication channel, $\D$ computes
	$\widetilde{X}^{m}$ in the following way.
        From (\ref{eqn:aSdzx}), we have that 
        the decoder $\D$ can obtain 
        $\widetilde{X}^{m}$ $=\phi^{(n)}({\rvcx})$
        by subtracting $\widetilde{K}^{m}=\varphi^{(n)}({\rvck})$ 
        from $\widetilde{C}^{m}$. 
	Finally, $\D$ outputs $\hrvcx$
	by applying the decoder 
        $\psi^{(n)}$ to 
	$\widetilde{X}^{m}$ as follows:
	\begin{align}
		\hrvcx &=\psi^{(n)}(\widetilde{X}^{m}) 
	 	       =\psi^{(n)}(\phi^{(n)}({\rvcx})). 
                  \label{eq:source_estimation}
	\end{align}
\end{enumerate}
Our privacy amplified system described above is illustrated 
in Fig. \ref{fig:solution}.

\noindent
\underline{\it On Reliability:}
From the description of our system in the previous section,
the decoding process in our system above is successful if 
$\widehat{X}^{n}=X^{n}$ holds.
Combining this and (\ref{eq:source_estimation}), 
it is clear that the decoding error probability 
$p_{\rm e}$ is as follows:
\begin{align}
p_{\rm e}=& p_{\rm e}(\phi^{(n)},\psi^{(n)}|{p}_{X}^n)
\defeq \Pr[\psi^{(n)}(\phi^{(n)}({\rvcx}))\neq {\rvcx}]. 
\notag
\end{align}

\noindent
\underline{\it On Security:} 
Set $M_{\cal A}^{(n)}=\varphi_{\cal A}^{(n)}(Z^n)$. 
The adversary 
${\cal A}$ tries to estimate ${\rvcx} 
\in \mathcal{X}^n$ from
\begin{align*}
&(\widetilde{C}^{m},M_{\cal A}^{(n)})
=(\varphi^{(n)}({\rvcx} \oplus {\rvck}), M_{\cal A}^{(n)}) 
\in \mathcal{X}^{m} \times \mathcal{M}_{\cal A}^{(n)}.
\end{align*}
We assume that the adversary ${\cal A}$ knows $(A,b^n)$ 
defining the affine encoder $\varphi^{(n)}$. 
The information leakage $\Delta^{(n)}$ on $X^n$ from 
$(\widetilde{C}^{m},M_{\cal A}^{(n)})$
is measured by the mutual information 
between $X^n$ and $(\widetilde{C}^{m},$ 
$M_{\cal A}^{(n)})$.
This quantity is formally defined by 
\begin{align*}
&\Delta^{(n)}=
\Delta^{(n)}(\varphi^{(n)},\varphi_{\cal A}^{(n)}|
{p}_{X}^n, {p}_{K}^n,W^n)
\\
&\defeq I(X^n;\widetilde{C}^{m},M_{\cal A}^{(n)})  
       = I(X^n;\varphi^{(n)}({\rvcx}\oplus {\rvck}),M_{\cal A}^{(n)}).  
\end{align*}

\noindent
\underline{\it Reliable and Secure Framework:} \ 

\begin{definition}
        A quantity $R$ is achievable under $R_{\cal A}$ $>0$
        for the system $\mathsf{Sys}$ if there exists 
        a sequence $\{(\varphi^{(n)},$ $\psi^{(n)})\}_{n \geq 1}$
        such that 
        $\forall \epsilon>0$, $\exists n_0=n_0(\epsilon) \in\mathbb{N}_0$, 
	$\forall n\geq n_0$, we have
	\begin{align*}
        &\frac {1}{n} \log |{\cal X}^{m}| 
        = \frac {m}{n} \log |{\cal X}|
        \leq R+\epsilon,\\  				
        &p_{\rm e}(\phi^{(n)},\psi^{(n)}|{p}_{X}^n) \leq \epsilon
	\end{align*}
	and for any eavesdropper $\A$ with $\varphi_{\A}$ satisfying 
        $\varphi_{\cal A}^{(n)}
        \in {\cal F}_{\cal A}^{(n)}(R_{\cal A}+\epsilon)$, we have 
	\begin{align*}
        \Delta^{(n)}(\varphi^{(n)},\varphi_{\cal A}^{(n)}
        |{p}_{X}^n,{p}_{K}^n,W^n)
	\leq \epsilon.
	\end{align*}
\end{definition}

\begin{definition}{\bf (Reliable and Secure Rate Region)}
        Let $\mathcal{R}_{\mathsf{Sys}}({p}_{X},$ ${p}_{K},W)$
	denote the set of all $(R_{\cal A},R)$ such that
        $R$ is achievable under $R_{\cal A}$. 
        We call $\mathcal{R}_{\mathsf{Sys}}({p}_{X},{p}_{K},$ $W)$
        the \emph{\bf reliable and secure rate} region. 
\end{definition}

\begin{definition}
        A triple $(R,E,F)$ is achievable under \\ 
        $R_{\cal A}>0$
        for the system $\mathsf{Sys}$ if there exists 
        a sequence $\{(\varphi^{(n)},$ $\psi^{(n)})\}_{n \geq 1}$
        such that $\forall \epsilon>0$, 
        $\exists n_0=n_0(\epsilon)\in\mathbb{N}_0$, 
	$\forall n$ $\geq n_0$, we have
	\begin{align*}
        &\frac {1}{n} \log |{\cal X}^{m}|=\frac {m}{n}
         \log |{\cal X}| \leq R+\epsilon, 
        \\
        & p_{\rm e}(\phi^{(n)},\psi^{(n)}|{p}_{X}^n) 
        \leq {\ExP}^{-n(E-\epsilon)},
	\end{align*}
	and for any eavesdropper $\A$ with $\varphi_{\A}$
        satisfying $\varphi_{\cal A}^{(n)} 
        \in {\cal F}_{\cal A}^{(n)}(R_{\cal A}+\epsilon)$, 
        we have 
	\begin{align*}
	\Delta^{(n)}(\varphi^{(n)},\varphi_{\cal A}^{(n)}
        |{p}_{X}^n,{p}_{K}^n,W^n)
        \leq {\ExP}^{-n(F-\epsilon)}.
	\end{align*}
\end{definition}

\begin{definition}{\bf (Rate Reliability and Security Region)}
        Let $\mathcal{D}_{\mathsf{Sys}}({p}_{X},$ ${p}_{K},W)$
	denote the set of all $(R_{\cal A},R,E,F)$ such that
        $(R,E,F)$ is achievable under $R_{\cal A}$. 
        We call $\mathcal{D}_{\mathsf{Sys}}({p}_{X},$ ${p}_{K},W)$
        the \emph{\bf rate reliability and security} region.
\end{definition}

\section{Main Results}

In this section we state our main results. To describe our results we define
several functions and sets. Let $U$ be an auxiliary random 
variable taking values in a finite set ${\cal U}$.  
We assume that the joint distribution of $(U,Z,K)$ is 
$$ 
p_{U{Z}{K}}(u,z,k)=p_{U}(u)
p_{{Z}|U}(z|u)p_{K|Z}(k|z).
$$
The above condition is equivalent to $U \markov Z \leftrightarrow K$. 
Define the set of probability distribution $p=p_{UZK}$
by
\begin{align*}
&{\cal P}(p_{K},W)
\defeq 
\{p_{UZK}: \pa {\cal U} \pa \leq \pa {\cal Z} \pa+1,
U \markov  Z \markov K \}.
\end{align*}
Set
\begin{align*}
{\cal R}(p)
\defeq &
\ba[t]{l}
\{(R_{\cal A},R):R_{\cal A},R \geq 0,
\vSpa\\
\: R_{\cal A}\geq I_{\empty}({Z};{U}), R \geq H_{\empty}({K}|{U})\},
\ea
\\
{\cal R}(p_{K},W)\defeq& \bigcup_{p \in {\cal P}(p_{K},W)}
{\cal R}(p).
\end{align*}
\newcommand{\ZaSS}{
}
{
We can show that the region ${\cal R}(p_K,W)$ satisfies the 
following property.
\begin{property}\label{pr:pro0}  
$\quad$
\begin{itemize}
\item[a)] 
The region ${\cal R}(p_{K},W)$ is a closed convex subset 
of $\mathbb{R}_{+}^2:=\{ R_{\cal A}\geq 0,R\geq 0 \}$.
\item[b)] For any $(p_{K},W)$, we have 
\beq
\min_{(R_{\cal A},R)\in {\cal R}(p_{K},W)}(R_{\cal A}+R)=H_{\empty}(K).
\label{eqn:SdEEE}
\eeq
The minimum is attained by $(R_{\cal A},R)=(0,H_{\empty}(K$$))$. 
This result implies that 
\begin{align*}
 {\cal R}(p_{K},W) \subseteq &
\{(R_{\cal A},R): R_{\cal A}+R \geq H_{\empty}(K)\} 
\cap \mathbb{R}_{+}^2.
\end{align*} 
Furthermore, the point $(0,H_{\empty}(K))$ always belongs to 
${\cal R}(p_{K},W)$. 
\end{itemize}
\end{property}

Property \ref{pr:pro0} part a) is a well known property.
Proof of Property \ref{pr:pro0} part b) is easy. 
Proofs of Property \ref{pr:pro0} parts a) and b) are omitted. 
}

Our result on $\mathcal{R}_{\mathsf{Sys}}({p}_{X},$ ${p}_{K},W)$
is the following: 
\begin{theorem} 
\begin{align*}
 \mathcal{R}^{\rm (in)}_{\mathsf{Sys}}({p}_{X},{p}_{K},W)
  \defeq &\{R \geq H_{\empty}(X)\} 
\cap {\rm cl}\left[{\cal R}^{c}(p_{K},W)\right] 
\\
\subseteq & \mathcal{R}_{\mathsf{Sys}}({p}_{X},{p}_{K},W),
\end{align*}
\end{theorem}
where ${\rm cl}\left[{\cal R}^{\rm c}(p_{K},W)\right]$ stands for 
the closure of the complement of ${\cal R}(p_{K},W)$.

This theorem is proved by several techniques Watanabe and Oohama
developed for establishing the direct part of privacy amplification 
theorem for bounded storage eavesdropper posed by them. We omit the detail.
The privacy amplification for bounded storage eavesdropper
has some interesting duality with the one helper source 
coding problem posed and investigated by Ashlswede and K\"orner 
\cite{ahlswede:75} and Wyner \cite{wyner:75c}. 

We next define several quantities to state a result on 
$\mathcal{R}_{\mathsf{Sys}}({p}_{X},{p}_{K},W)$. 
We first define a function related to an exponential upper bound of 
$p_{\rm e}(\phi^{(n)},\psi^{(n)}|{p}_{X}^n)$. 
Let $\overline{X}$ be an arbitrary random variable
over $\mathcal{X}$ and has a probability distribution 
$p_{\overline{X}}$. Let $\mathcal{P}(\mathcal{\cal X})$ 
denote the set of all probability distributions on 
$\mathcal{X}$. 
For $ R \geq 0$ and 
$p_{X} \in $ $\mathcal{P}(\mathcal{\cal X})$, 
we define the following function:
\begin{align*}
	E(R|p_{X}) &:{=}
	\min_{ p_{\overline{X}} \in 
	\mathcal{P}(\mathcal{\cal X})}
	\{[R- H(\overline{X})]^{+}
		+D(p_{\overline{X}}||p_X)\}.
\end{align*}
We next define a function related to an exponential 
upper bound of 
$\Delta^{(n)}(\varphi^{(n)},\varphi_{\A}^{(n)}$ 
$|{p}_{X}^n,{p}_{K}^n,W^n)$. 
Set
\begin{align*}
{\cal Q}(p_{K|{Z}}) \defeq & \{q=q_{U{Z}K}: 
\pa {\cal U} \pa \leq \pa {\cal {Z}} \pa,
{U} \markov {{Z}} \markov {K}, 
\\
& p_{K|{Z}}=q_{K|{Z}}\}.
\end{align*}
For $(\mu,\alpha) \in [0,1]^2$, and 
for $q=q_{U{Z}{K}}\in {\cal Q}(p_{K|{Z}})$, 
define 
\begin{align*}
& \omega_{q|p_{Z}}^{(\mu,\alpha)}({z},k|u)
\\
& \defeq 
 \bar{\alpha} \log \frac{q_{{Z}}({z})}{p_{{Z}}({z})}
+ \alpha \left[
  {\mu}\log \frac{q_{{Z}|U}({z}|u)}{p_{{Z}}({z})}\right.
\left. +\bar{\mu}\log\frac{1}{q_{K|U}(k|u)}
\right],
\\
& \Omega^{(\mu,\alpha)}(q|p_{Z})
\defeq 
-\log {\rm E}_{q}
\left[\exp\left\{-\omega^{(\mu,\alpha)}_{q|p_{Z}}({Z},K|U)\right\}\right],
\\
&\Omega^{(\mu,\alpha)}(p_{K},W)
\defeq 
\min_{\scs 
   \atop{\scs 
    q \in {{\cal Q}}(p_{K|{Z}})
   }
}
\Omega^{(\mu,\alpha)}(q|p_{Z}),
\\
& F^{(\mu,\alpha)}({\mu}R_{\cal A}+\bar{\mu} R_{\empty} |p_{K},W)
\\
&\defeq \frac{\Omega^{(\mu,\alpha)}(p_{K},W)
-\alpha({\mu}R_{\cal A} + \bar{\mu} R_{\empty})}
{2+\alpha \bar{\mu}},
\\
& F(R_{\cal A},R_{\empty}|p_{K},W)
\defeq \sup_{ (\mu,\alpha)\in [0,1]^2}
F^{(\mu,\alpha)}({\mu}R_{\cal A}+ \bar{\mu} R_{\empty}|p_{K},W).
\end{align*}
We next define a function serving as a lower 
bound of $F(R_{\cal A},R_{\empty}|p_{K},W)$. 
For each $p_{U{Z}K}\in {\cal P}_{\rm sh}(p_{K},W)$, 
define 
\begin{align*}
& \tilde{\omega}_{p}^{(\mu)}({z},k|u)
\defeq {\mu}\log \frac{p_{{Z}|U}({z}|u)}{p_{{Z}}({z})}
  +\bar{\mu}\log \frac{1}{p_{K|U}(K|U)},
\\
& \tilde{\Omega}^{(\mu,\lambda)}(p)
\defeq 
-\log 
{\rm E}_{p}
\left[\exp\left\{-\lambda
\tilde{\omega}_p^{(\mu)}({Z},K|U)\right\}\right].
\end{align*}
Furthermore, set
\begin{align*}
& \tilde{\Omega}^{(\mu,\lambda)}(p_{K},W)
 \defeq \min_{\scs \atop{\scs 
  p\in {{\cal P}_{\rm sh}(p_{K},W)}}}
\tilde{\Omega}^{(\mu,\lambda)}(p),
\\
& \tilde{F}^{(\mu,\lambda)}
  ({\mu}R_{\cal A}+\bar{\mu}R_{\empty}|p_{K},W) 
\\
&\defeq 
\frac{\tilde{\Omega}^{(\mu,\lambda)}(p_{K},W)
-\lambda({\mu}R_{\cal A}+R_{\empty})}{2+\lambda(5-{\mu})},
\\
& \tilde{F}(R_{\cal A},R_{\empty}|p_{K},W)
\defeq \sup_{\scs \lambda \geq 0, \atop{\scs \mu \in [0,1]}} 
\tilde{F}^{(\mu,\lambda)}({\mu} R_{\cal A} + \bar{\mu}R_{\empty}|p_{K},W).
\end{align*}
We can show that the above functions satisfy 
the following property. 
\begin{property}\label{pr:pro1}  
$\quad$
\begin{itemize}
\item[a)] 
The cardinality bound 
$|{\cal U}|\leq |{\cal {Z}}|$ in ${\cal Q}(p_{K|{Z}})$
is sufficient to describe the quantity
$\Omega^{(\mu,\beta,\alpha)}(p_{K},W)$. 
Furthermore, the cardinality bound 
$|{\cal U}|\leq |{\cal {Z}}|$ in ${\cal P}_{\rm sh}(p_{K},W)$
is sufficient to describe the quantity
$\tilde{\Omega}^{(\mu,\lambda)}(p_{K},W)$. 

\item[b)] For any $R_{\cal A},R_{\empty}\geq0$, we have 
\beqno
& &F(R_{\cal A},R_{\empty}|p_{K},W)\geq 
\tilde{F}(R_{\cal A},R_{\empty}|p_{K},W).
\eeqno

\item[c)] For any $p=p_{UZK} \in {\cal P}_{\rm sh}(p_{Z},W)$
and any $(\mu,\lambda)\in [0,$ $1]^2$, we have 
\beq
0\leq \tilde{\Omega}^{(\mu,\lambda)}(p) 
 \leq \prmtA \log |{\cal Z}|
    + \prmtB \log |{\cal K}|.
\label{eqn:Asddx}
\eeq

\item[d)] Fix any $p=p_{U{Z}K}\in {\cal P}_{\rm sh}(p_{K},W)$ and $\mu\in [0,1]$. 
For $\lambda \in [0,1]$, we define a probability distribution 
$p^{(\lambda)} = p_{U{Z}K}^{(\lambda)}$ by
\begin{align*} 
& p^{(\lambda)}(u,{z},k)
\defeq
\frac{
p(u,{z},k)
\exp\left\{
-\lambda \tilde{\omega}^{(\mu)}_p({z},k|u)
\right\}
}
{{\rm E}_{p}
\left[\exp\left\{-\lambda
\tilde{\omega}^{(\mu)}_p({Z},K|U)\right\}\right]}.
\end{align*}
Then for $\lambda \in [0,1/2]$, 
$\tilde{\Omega}^{(\mu,\lambda)}(p)$ is 
twice differentiable.
Furthermore, for $\lambda \in [0,1/2]$, we have
\beqno
& & \frac{\rm d}{{\rm d}\lambda}
\tilde{\Omega}^{(\mu,\lambda)}(p)
={\rm E}_{p^{(\lambda)}}
\left[\tilde{\omega}_p^{(\mu)}({Z},K|U)\right],
\\
& & \frac{\rm d^2}{{\rm d}\lambda^2} 
\tilde{\Omega}^{(\mu,\lambda)}(p)
=-{\rm Var}_{p^{(\lambda)}}
\left[\tilde{\omega}^{(\mu)}_{p}({Z},K|U)\right].
\eeqno
The second equality implies that 
$\tilde{\Omega}^{(\mu,\lambda)}(p|p_{K}$$,W)$ 
is a concave function of $\lambda\geq0$. 
\item[e)] 
For $(\mu,\lambda)\in [0,1] \times [0,1/2]$, define 
\begin{align*}
& \rho^{(\mu,\lambda)}(p_{K},W)
\\
&\defeq {\max_{\scs (\nu, p) \in [0,\lambda] 
    \atop{\scs \times {\cal P}_{\rm sh}(p_{K},W):
        \atop{\scs \tilde{\Omega}^{(\mu,\lambda)}(p) 
             \atop{\scs 
             =\tilde{\Omega}^{(\mu,\lambda)}(p_{K},W)
             }}}}
}
{\rm Var}_{{p^{(\nu)}}}\left[\tilde{\omega}^{(\mu)}_{p}({Z},K|U)\right],
\end{align*}
and set
\begin{align*}
& \rho =\rho(p_{K},W)
\defeq \max_{(\mu,\lambda)\in [0,1] \times [0,1/2]}
\rho^{(\mu,\lambda)}(p_{K},W).
\end{align*}
Then we have $\rho(p_{K},W)<\infty$. Furthermore, 
for any $(\mu,\lambda) \in [0,1]\times[0,1/2]$, we have
$$
\tilde{\Omega}^{(\mu,\lambda)}(p_{K},W) 
\geq \lambda R^{(\mu)}(p_{K},W)
-\frac{\lambda^2}{2}\rho(p_{K},W).
$$
\item[f)] For every $\tau \in(0,(1/2)\rho(p_{K},W))$,
the condition 
$(R_{\cal A},$ $R_{\empty}+\tau) \notin {\cal R}(p_{K},W)$
implies 
\begin{align*}
& \tilde{F}(R_{\cal A},R_{\empty}|p_{K},W)
> \ts \frac{\rho(p_{K},W)}{4} \cdot g^2
\left({\ts \frac{\tau}{\rho(p_{K},W)}}\right)>0,
\end{align*}
where $g$ is the inverse function of 
$\vartheta(a) \defeq a+(5/4)a^2, a \geq 0$.
\end{itemize}
\end{property}

Proof of this property is found in 
Oohama \cite{oohama2015exponent}(extended version).
Our main result is as follows. 
\begin{theorem}\label{Th:mainth2}{
\rm For any $R_{\cal A}, R>0$, and 
any $(p_K,W)$, there exists a sequence of mappings 
$\{(\varphi^{(n)}, \psi^{(n)}) \}_{n=1}^{\infty}$
such that for any $p_X$ with $(R_{\cal A},R) \in$
$\mathcal{R}_{\sf Sys}(p_{X},p_{K},W)$,
we have 
\begin{align}
& \frac {1}{n} 
\log |{\cal X}^{m}|= \frac {m}{n} \log |{\cal X}|\leq R,
\notag\\
& p_{\rm e}(\phi^{(n)},\psi^{(n)}|p_{X}^n) \leq 
{\ExP}^{-n[E(R|p_{X})-\delta_{1,n}]}
\label{eqn:mainThErrB}
\end{align}
and for any eavesdropper $\A$ with $\varphi_{\A}$ satisfying
$\varphi_{\A}^{(n)} \in {\cal F}_{\A}^{(n)}(R_{\A})$, we have
\begin{align}
& \Delta^{(n)}(\varphi^{(n)},\varphi_{\A}^{(n)}|p_{X}^n,p_{K}^n,W^n)
\notag\\
&\leq {\ExP}^{-n[F(R_{\A},R|p_{K},W)-\delta_{2,n}]}, 
\label{eqn:mainThSecB}
	\end{align}
where $\delta_{i,n},i=1,2$ are defined by
\begin{align*}
&\delta_{1,n}:=
\frac{1}{n}\log\left[
{\ExP}(n+1)^{2|{\cal X}|}\{(n+1)^{|{\cal X}|}+1\} 
\right],
\\
&
\delta_{2,n}
:=\frac{1}{n} \log 
\left[5nR \{(n+1)^{|{\cal X}|}+1\}\right].
\end{align*}
Note that for $i=1,2$, $\delta_{i,n} \to 0$ as $n\to \infty$. 
}
\end{theorem}

This theorem is proved by a coupling of two techniques. One is a technique 
Watanabe and Oohama \cite{watanabe2012privacy} developed for establishing 
the direct part of privacy amplification 
theorem for bounded storage eavesdropper posed by them.
The other is a technique Oohama \cite{oohama2015exponent} developed 
for establishing exponential strong converse theorem 
for the one helper source coding problem.
The functions $E(R|p_{X})$ and $F(R_{\A},R|p_{K},W)$ 
take positive values if and only 
if $(R_{\A},R)$ belongs to the set
\begin{align*}
&\{R > H(X)\} \cap {\cal R}^{\rm c}(p_{K},W)
\defeq {\rm int}
\left[
{\cal R}_{\rm Sys}^{\rm (in)}
(p_{X},p_{K},W)
\right].
\end{align*}
Here ${\rm int}[{\cal R}]$ stands for the set of 
inner points of ${\cal R}$. 
Thus, by Theorem \ref{Th:mainth2}, 
under 
$$
(R_{\A},R) \in {\rm int} 
\left[{\cal R}_{\rm Sys}^{\rm (in)}(p_{X},p_{K},W)
\right],
$$  
we have the followings: 
\begin{itemize}
\item On the reliability, 
$p_{\rm e}(\phi^{(n)},\psi^{(n)}|p_{X}^n)$  
goes to zero exponentially as $n$ tends to infinity, and its 
exponent is lower bounded by the function $E(R|p_{X})$.  
\item On the security, for any $\varphi_{\cal A}$ satisfying
$\varphi_{\cal A}^{(n)}\in$ ${\cal F}_{\cal A}^{(n)}(R_{\cal A})$, 
the information leakage
$
\Delta^{(n)}(\varphi^{(n)},\varphi_{\A}^{(n)}$ $|p_{X}^n,p_{K}^n,W^n) 
$
on $X^n$ goes to zero exponentially as $n$ tends to infinity, and 
its exponent is lower bounded by the function $F(R_{\A},R|p_{K},W)$.
\item The code that attains the exponent functions 
$E($ $R|p_{X})$ is the universal code that depends only on $R$ 
not on the value of the distribution $p_{X}$. 
\end{itemize}
Define
\begin{align*}
&{\cal D}_{\rm Sys}^{\rm (in)}(p_{X},p_{K},W)
\\
&:=\{(R_1,R_2,E(R|p_{X}),F(R_{\A},R|p_{K})):
\notag\\
&\quad (R_1,R_2)\in {\cal R}_{\sf Sys}^{\rm (in)}(p_X,p_{K},W)\}.
\end{align*}

From Theorem \ref{Th:mainth2}, we immediately obtain 
the following corollary. 
\begin{corollary}
\begin{align*}
{\cal D}_{\rm Sys}^{\rm (in)}(p_{X},p_{K},W)
\subseteq {\cal D}_{\rm Sys}(p_{X},p_{K},W).
\end{align*}
\end{corollary}

\begin{figure}[t]
\centering
\includegraphics[width=0.36\textwidth]{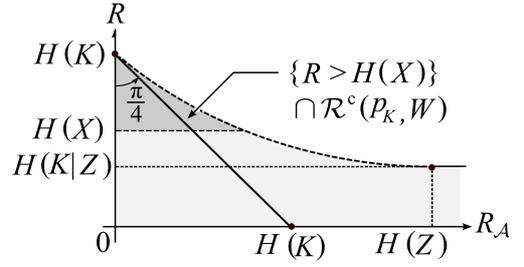}
\caption{
The inner bound 
${\rm int}[{R}_{\rm Sys}^{\rm (in)}(p_{X},p_K,W)]$
of the reliable and secure rate region 
${\cal R}_{\rm Sys}(p_{X},p_K$ $W)$.
}
\label{fig:admissible}
\end{figure}

A typical shape of $\{R>H(X)\}\cap \mathcal{R}(p_{K},W)$
is shown in Fig. \ref{fig:admissible}. 

\section{Proofs of the Results}

In this section we prove Theorem \ref{Th:mainth2}. 

\subsection{Types of Sequences and Their Properties}

In this subsection we prepare basic results on the types.  
Those results are basic tools for our analysis of 
several bounds related to error provability 
of decoding or security.
\begin{definition}{\rm
For any $n$-sequence ${\vcxone}=x_{1}x_{2}\cdots $ 
$x_{n}\in {\calVarX}^{n}$, 
$n(x|{\vcxone})$ denotes the number of $t$ such that $x_{t}=x$.  
The relative frequency $\left\{n(x|\vcxone)/n\right\}_{x\in {\cal X}}$ 
of the components of ${\vcxone}$ is called the type of ${\vcxone}$ 
denoted by $P_{\lvcxone}$.  The set that consists of all 
the types on ${\cal X}$ is denoted by ${\cal P}_{n}({\cal X})$. 
Let $\tX$ denote an arbitrary random variable whose distribution 
$P_{\tX}$ belongs to ${\cal P}_{n}({\cal X})$. 
For $p_{\tX}\in {\cal P}_{n}({\cal X})$, set 
$
T_{\tX}^n := \left\{{\vcxone}:\, P_{{\lvcxone}}=p_{\tX}\right\}.
$
}

\end{definition}
For set of types and joint types the following lemma holds. 
For the detail of the proof see Csisz\'ar 
and K\"orner \cite{csiszar-korner:11}.
\begin{lemma}\label{lem:Lem1}{\rm 
$\quad$

\begin{itemize}
\item[a)]
   $\begin{array}[t]{l} 
   |{\cal P}_{n}({\calVarX})|\leq(n+1)^{|{\calVarX}|}.  
   \end{array}$
  
\item[b)] For $P_{\tX}\in {\cal P}_{n}(\calVarX)$,
\begin{align*}
 (n+1)^{-{|\calVarX|}}{\ExP}^{nH(\Dist{\tX})}
&   \leq |T^n_{\tX}|\leq {\ExP}^{nH(\Dist{\tX})}.
\end{align*}
\item[c)] For ${\vcxone} \in T^n_{\tX}$, 
\begin{align*}
p_{X}^n({\vcxone})
&={\ExP}^{-n[H(\Dist{\tX})+D(p_{\Dist{\tX}}||p_X)]}.
\end{align*}
\end{itemize}
}
\end{lemma}

By Lemma \ref{lem:Lem1} parts b) and c), we immediately 
obtain the following lemma: 
\begin{lemma}\label{lem:Lem1.4}{\rm $\quad$
For $p_{\tX}\in {\cal P}_{n}({\calVarX})$,  
\begin{align*}
& p_{X}^n(T_{\tX }^n)
\leq {\ExP}^{-nD(p_{\Dist{\tX}}||p_{X})}.
\end{align*}
}
\end{lemma}

\subsection{Upper Bounds of 
$p_{\rm e}(\phi^{(n)},\psi^{(n)}|p_{X}^n)$ and 
$\Delta_n(\varphi^{(n)},\varphi^{(n)}_{\cal A}$ 
$|p_{X}^n, p_{K}^n, W^n)$}

In this subsection we evaluate upper bounds of 
$p_{\rm e}($ $\phi^{(n)},\psi^{(n)}| p_{X}^n)$ and 
$\Delta_n (\varphi^{(n)}, \varphi^{(n)}_{\cal A}$$|p_{X}^n,p_{K}^n,W^n)$. 
For $p_{\rm e}(\phi^{(n)}$ $,\psi^{(n)}| p_{X}^n)$, we derive 
an upper bound which can be characterized with a quantity 
depending on $(\phi^{(n)},\psi^{(n)})$ and type 
$P_{\lvcxone}$ of sequences $x^n \in {\cal X}^n $.
We first evaluate $p_{\rm e}(\phi^{(n)}, \psi^{(n)}|p_{X}^n)$.   
For ${\vcxone} \in {\cal X}^{n}$ and 
$p_{\overline{X}}
 \in{\cal P}_{n}({\cal X})$ we define the following functions.
\begin{align*}
\Xi_{{\lvcxone}}(\phi^{(n)},\psi^{(n)})
&\defeq\left\{
\begin{array}{ccl}
1&\qquad&\mbox{if}\quad
         \psi^{(n)}\bigl(\phi^{(n)}({\vcxone})\bigr)
	 \neq {\vcxone},
	 \vspace{0.2cm}\\
0&\qquad&\mbox{otherwise,}
\end{array}
\right.\\
\displaystyle \Xi_{\overline{X}}
(\phi^{(n)},\psi^{(n)})
&\defeq \frac{1}{|T_{\overline{X}}^{n}|}
\sum_{{\lvcxone} \in T_{\overline{X}}^{n}}
\Xi_{{\lvcxone}}(\phi^{(n)},\psi^{(n)}).
\end{align*}
Then we have the following lemma. 
\begin{lemma} 
\label{lem:ErBound}
In the proposed system, for any pair of 
$(\phi^{(n)},$ $\psi^{(n)})$, we have
\begin{align}
& p_{\rm e}(\phi^{(n)},\psi^{(n)}| p_{X}^n)
\notag\\
& \leq \sum_{ p_{\overline{X}}
     \in {\cal P}_n({\cal X}) }
  \Xi_{ \overline{X} }(\phi^{(n)},\psi^{(n)})
  {\ExP}^{-nD(p_{ \overline{X} }||p_{X})}.
\label{eqn:ErrBound}
\end{align}
\end{lemma}

\begin{IEEEproof} 
We have the following chain of inequalities:
\begin{align*}
&p_{\rm e} (\phi^{(n)}, \psi^{(n)} |p_{X}^n)
\\
&\MEq{a} \sum_{ p_{\overline{X}} 
\in {\cal P}_n({\cal X})}
\sum_{{\lvcxone}\in T^n_{ \overline{X}} }
\Xi_{\lvcxone}(\phi^{(n)},\psi^{(n)})
p_{X}^n({\vcxone})
\\
&= \sum_{ p_ {\overline{X}} \in {\cal P}_n({\cal X})}
\frac{1}{|T^n_{\overline{X}}|}
\sum_{{\lvcxone}\in T^n_{\overline{X}}}
\Xi_{{\lvcxone} }(\phi^{(n)},\psi^{(n)})
|T^n_{\overline{X}}|p_{X}^n({\vcxone})
\\
&\MEq{b}\sum_{p_{ \overline{X}} \in {\cal P}_n({\cal X})}
\frac{1}{|T^n_{\overline{X}}|}
\sum_{\lvcxone \in T^n_{\overline{X}}}
\Xi_{ \lvcxone }(\phi^{(n)},\psi^{(n)})
p_{X}^n(T^n_{ \overline{X} })
\\
&\MEq{c}\sum_{p_{\overline{X}} 
\in {\cal P}_n({\cal X})}
\Xi_{\overline{X}}
(\phi^{(n)},\psi^{(n)})
p_{X}^n(T^n_{\overline{X}})
\\
&
\MLeq{d}\sum_{p_{\overline{X}}
\in {\cal P}_n({\cal X})}
\Xi_{\overline{X}}(\phi^{(n)},\psi^{(n)})
{\ExP}^{-nD(p_{\overline{X}}||p_{X})}.
\end{align*}
Step (a) follows from the definition of 
$\Xi_{{\lvcxone}}(\phi^{(n)},\psi^{(n)})$.
Step (b) follows from that the probabilities 
$p_{X}^n(\vcxone)$ for $ \vcxone \in T^n_{\overline{X}}$
take an identical value. 
Step (c) follows from the definition of 
$\Xi_{\overline{X}}(\phi^{(n)},\psi^{(n)})$.
Step (d) follows from lemma \ref{lem:Lem1.4}. 
\end{IEEEproof}

We next discuss upper bounds of 
$$
\Delta_{n}(\varphi^{(n)},\varphi_{\cal A}^{(n)}|p_{X}^n, p_K^n, W^n)
=I(\widetilde{C}^{m}, M_{\cal A}^{(n)};X^{n}), 
$$
On an upper bound  of 
$I(\widetilde{C}^{m}, M_{\cal A}^{(n)};X^{n})$, 
we have the following lemma.
\begin{lemma}\label{lem:MIandDiv}
\begin{align}
& I(\widetilde{C}^{m}, M_{\cal A}^{(n)};X^{n}) 
\leq 
\left.\left.\left.\!\!\!
D\left(p_{\widetilde{K}^m|M_{{\cal A}}^{(n)}}
\right|\right|
p_{V^m} \right| p_{M_{\cal A}^{(n)}}\right),
\label{eqn:oohama}
\end{align}
where $p_{V^m}$ represents the uniform distribution
over $\mathcal{X}^{m}$. 
\end{lemma}

\begin{IEEEproof}
We have the following chain of inequalities: 
\begin{align*}
& I(\widetilde{C}^{m}, M_{\cal A}^{(n)};{\rvcxone})
\MEq{a}I(\widetilde{C}^{m};{\rvcxone}| M_{\cal A}^{(n)})
\\
&\leq \log |{\cal X}^{m}|
 -H(\widetilde{C}^{m}|{\rvcxone}, M_{\cal A}^{(n)})
\\
&\MEq{b} \log |{\cal X}^{m}|
        -H(\widetilde{K}^{m}|{\rvcxone},M_{\cal A}^{(n)})
\\
& \MEq{c} \log |{\cal X}^{m}|
         -H(\widetilde{K}^{m}|M_{\cal A}^{(n)})
\\
&=\left.\left.\left. \!\!
D\left(p_{\widetilde{K}^m|M_{{\cal A}}^{(n)}}
\right|\right|
p_{V^m} \right| p_{M_{\cal A}^{(n)}}\right).
\end{align*}
Step (a) follows from ${\rvcxone} \perp M_{\cal A}^{(n)}$. 
Step (b) follows from 
$\widetilde{C}^{m}=\widetilde{K}^{m} 
    \oplus \widetilde{X}^{m}$ and $\widetilde{X}^{m}=\phi^{(n)}({\rvcx})$.
Step (c) follows from 
$(\widetilde{K}^{m},M_{\cal A}^{(n)}) \perp {\rvcxone}$. 
\end{IEEEproof}

\subsection{Random Coding Arguments}

We construct a pair of affine encoders 
$\varphi^{(n)}=(\varphi_1^{(n)},\varphi_{\ExP}^{(n)})$ 
using the random coding method. For the joint decoder 
$\psi^{(n)}$, we propose the minimum entropy decoder 
used in Csisz\'{a}r \cite{csiszar:82} and 
Oohama and Han \cite{oohama:94}.
 
\noindent
\underline{\it Random Construction of Affine Encoders:} \  
We first choose $m$ such that 
$$
m:=\left\lfloor\frac{nR}{\log |{\cal X}|}\right\rfloor,
$$
where $\lfloor a \rfloor$ stands for the integer part of $a$. 
It is obvious that  
$$
R-\frac{1}{n} \leq \frac{m}{n}\log |{\cal X}|\leq R. 
$$
By the definition (\ref{eq:homomorphica}) of 
$\phi^{(n)}$, we have that for ${\vcx} \in {\cal X}^n$,
\begin{align*}
& \phi^{(n)}({\vcx})={\vcx} A,
\end{align*}
where $A$ is a matrix with $n$ rows and $m$ columns.
By the definition (\ref{eq:homomorphic}) of 
$\varphi^{(n)}$, we have that for ${\vck} \in {\cal X}^n$,
\begin{align*}
& \varphi^{(n)}({\vck})={\vck} A+b^{m},
\end{align*}
where $b^{m}$ is a vector with $m$ columns.
Entries of $A$ and $b^{m}$ 
are from the field of ${\cal X}$. Those entries 
are selected at random, 
independently of each other and 
with uniform distribution.
Randomly constructed  
linear encoder $\phi^{(n)}$ and affine encoder 
$\varphi^{(n)}$ have three properties shown 
in the following lemma.
\newcommand{\LemmaAffineB}{
\begin{lemma}[Properties of Linear/Affine Encoders]
\label{lem:good_set}
$\quad$
\begin{itemize}
\item[a)] 
For any ${\vcx}, {\vcv} \in {\cal X}^n$ with 
         ${\vcx} \neq {\vcv}$, we have
	\begin{align} 
        &\Pr[\phi^{(n)}({\vcx})=
        \phi^{(n)}({\vcv})]=\Pr[({\vcx} \ominus {\vcv}) A =0^{m}]
        \notag \\
         &=|\mathcal{X}|^{-m}.
	\end{align}

\item[b)] For any ${\vcs} \in {\cal X}^n$, 
         and for any $\widetilde{s}^{m}\in {\cal X}^{m}$, we have
	\begin{align} 
	  &\Pr[\varphi^{(n)}({\vcs})=\widetilde{s}^{m}]=
          \Pr[\vcs A \oplus b^{m}=\widetilde{s}^{m}]
         \notag\\
         &=|\mathcal{X}|^{-m}.
	\end{align}
\item[c)] 
For any ${\vcs}, {\vct} \in {\cal X}^n$ 
with ${\vcs} \neq {\vct}$, 
and for any $\widetilde{s}^{m}
\in {\cal X}^{m}$, we have
	\begin{align} 
	&\Pr[\varphi^{(n)}({\vcs})=
            \varphi^{(n)}({\vct})=\widetilde{s}^{m}]
        \notag\\
        &= \Pr[{\vcs} A \oplus b^{m}=
               {\vct} A \oplus b^{m}=\widetilde{s}^{m}]
	\notag\\  
	& = |\mathcal{X}|^{-2m}.
	\end{align}
\end{itemize}
\end{lemma}
}
\LemmaAffineB

Proof of this lemma is given in Appendix \ref{apd:ProofLemAA}. 
\newcommand{\ProofLemAA}{
\subsection{Proof of Lemma \ref{lem:good_set}}
\label{apd:ProofLemAA}

{\it Proof:}
Let $a_l^m$ be the $l$-th low vector of the matrix $A$.
For each $l=1,2,\cdots,n$, let $A_l^m \in{\cal X}^m$ 
be a random vector which represents 
the randomness of the choice of 
$a_l^m \in{\cal X}^m$. 
Let $B^m \in{\cal X}^m$ be a random vector which represent 
the randomness of the choice of $b^m \in{\cal X}^m$.
We first prove the part a).  
Without loss of generality we may assume 
$x_1 \neq v_1$. 
Under this assumption we have the following:
\begin{align}
& ({\vcx}\ominus \vcv ) A=0^m \Leftrightarrow 
 \sum_{l=1}^n(x_l \ominus v_l)a_l^m=0^m 
\notag\\
& \Leftrightarrow a_1^m=\sum_{l=2}^n 
\frac{v_l \ominus x_l}{x_1 \ominus v_1}a_l^m.
\label{eqn:Awff}
\end{align}
Computing $\Pr[\phi({\vcx})=\phi({\vcv})]$,
we have the following chain of equalities:
\begin{align*}
&\Pr[\phi({\vcx})=\phi({\vcv})]
=\Pr[(\vcy \ominus \vcw )A=0^m]
\\
& \MEq{a}\Pr\left[ a_1^m=\sum_{l=2}^n 
\frac{w_l\ominus y_l}{x_1 \ominus v_1}a_l^m \right]
\\
&\MEq{b}
\sum_{ \scs \left\{a_l^m \right\}_{l=2}^n 
        \atop{\scs 
        \in {\cal X}^{(n-1)m}
        }
    }   
\prod_{l=2}^n P_{A_l^m}(a_l^m)      
P_{A_1^m}\left(\sum_{l=2}^n 
\frac{w_l\ominus x_l}{y_1 \ominus v_1}a_l^m \right)
\\
& =|\mathcal{X}|^{-m}
 \sum_{ \scs \left\{a_l^m \right\}_{l=2}^n 
        \atop{
        \scs \in {\cal X}^{(n-1)m}
       }
    }   
\prod_{l=2}^n P_{A_l^m}(a_l^m)      
=|\mathcal{X}|^{-m}.
\end{align*} 
Step (a) follows from (\ref{eqn:Awff}). 
Step (b) follows from that $n$ random vecotors 
$A_l^m, l=1,2,\cdots,n$ are independent. 
We next prove the part b). We have 
the following:
\begin{align}
{\vcs} A \oplus b^{m}=\widetilde{s}^{m}
\Leftrightarrow 
b^m=\widetilde{s}^{m} \ominus 
\left\{ \sum_{l=1}^n s_la_l^m \right\}. 
\label{eqn:Awffb}
\end{align}
Computing $\Pr[{\vcs} A \oplus b^{m}=\widetilde{s}^{m}]$,
we have the following chain of equalities:
\begin{align*}
&\Pr[{\vcs} A \oplus b^{m}=\widetilde{s}^{m}]
\MEq{a}\Pr\left[b^m=\widetilde{s}^{m} \ominus 
 \left\{ \sum_{l=1}^n s_la_l^m\right\} 
\right]
\\
&\MEq{b} 
\sum_{ \scs \left\{a_l^m \right\}_{l=1}^n 
       \atop{\scs 
        \in {\cal X}^{nm}
       }
    }   
\prod_{l=1}^n P_{A_l^m}(a_l^m)      
P_{B^m}\left(
\widetilde{s}^{m} \ominus 
 \left\{ \sum_{l=1}^n s_la_l^m\right\}
\right)
\\
&=|\mathcal{X}|^{-m}
  \sum_{ \scs \left\{a_l^m \right\}_{l=1}^n 
        \atop{\scs 
        \in {\cal X}^{nm}
        }
    }   
\prod_{l=1}^n P_{A_l^m}(a_l^m)=|\mathcal{X}|^{-m}. 
\end{align*} 
Step (a) follows from (\ref{eqn:Awffb}). 
Step (b) follows from that 
$n$ random vectors $A_l^m, l=1,2, \cdots,n$ 
and $B^m$ are independent. 
We finally prove the part c). We first observe that
$
{\vcs} \neq {\vct} \Leftrightarrow 
$
is equivalent to 
$
s_i \neq t_i 
\mbox{ for some } i \in \{1,2,\cdots,n\}. 
$
Without loss of generality, we may assume that
$s_1 \neq t_1$. Under this assumption we have 
the following:
\newcommand{\Argument}{
\begin{align}
&\left.
\ba{l}
{\vcs} A \oplus b^{m}=\widetilde{s}^{m},
\\
{\vct} A \oplus b^{m}=\widetilde{t}^{m}
\ea
\right\}
\\
&\Leftrightarrow 
\left.
\ba{l}
({\vcs}\ominus {\vct})A=\widetilde{s}^m\ominus\widetilde{t}^m, 
\\
b^m=\ds \widetilde{s}^{m} \ominus 
\left\{\sum_{l=1}^n s_la_l^m \right\} 
\ea
\right\}
\notag\\
&\Leftrightarrow 
\left.
\ba{l}
\ds a_1^m=\sum_{l=2}^n 
   \frac{t_l \ominus s_l}{s_1 \ominus t_1} a_l^m
   \oplus 
\frac{\widetilde{s}^m \ominus \widetilde{t}^m}{s_1\ominus t_1},
\\
\ds b^m= \widetilde{s}^{m} \ominus 
\left\{ \sum_{l=1}^n s_l a_l^m \right\} 
\ea
\right\}
\Leftrightarrow 
\left.
\ba{l}
\ds a_1^m=\sum_{l=2}^n 
   \frac{t_l \ominus s_l}{s_1 \ominus t_1} a_l^m
   \oplus 
\frac{\widetilde{s}^m \ominus \widetilde{t}^m}{s_1\ominus t_1},
\\
\ds b^m=
\sum_{l=2}^n
\frac{t_1s_l\ominus s_1t_l}
{s_1\ominus t_1}a_l^m
\ominus \frac{t_1\widetilde{s}^m  \ominus s_1 \widetilde{t}^m }
{s_1\ominus t_1} 
\ea
\right\}.
\label{eqn:Awffc}
\end{align}
Then we have the following chain of equalities:
\begin{align*}
&\Pr[ {\vcs} A \oplus b^{m}=\widetilde{s}^{m}
\land {\vct} A \oplus b^{m}=\widetilde{t}^{m}]
\\
&\MEq{a}
\Pr\left[
a_1^m=\sum_{l=2}^n\frac{t_l \ominus s_l}{s_1 \ominus t_1} a_l^m
\oplus \frac{\widetilde{s}^m \ominus \widetilde{t}^m}{s_1\ominus t_1}
\land
b^m=
\sum_{l=2}^n
\frac{t_1s_l \ominus s_1t_l}
{s_1\ominus t_1}a_l^m
\ominus \frac{t_1\widetilde{s}^m  \ominus s_1 \widetilde{t}^m }
{s_1\ominus t_1} 
\right]
\\
&\MEq{b}
\sum_{ \scs \left\{ a_l^m \right \}_{l=2}^n 
      \atop{\scs
        \in {\cal X}^{(n-1)m}
      }
    }   
\prod_{l=2}^n P_{A_l^m}(a_l^m)      
P_{A_1^m}\left(\sum_{l=2}^n\frac{t_l \ominus s_l}{s_1 \ominus t_1} a_l^m
\oplus \frac{\widetilde{s}^m \ominus \widetilde{t}^m}{s_1\ominus t_1}
\right)
\\
&\qquad \times P_{B^m}
\left(
\sum_{l=2}^n
\frac{t_1s_l\ominus s_1t_l}
{s_1\ominus t_1}a_l^m
\ominus \frac{t_1\widetilde{s}^m  \ominus s_1 \widetilde{t}^m }
{s_1\ominus t_1} 
\right)
=|\mathcal{X}|^{-2m}
 \sum_{ \scs \left\{a_l^m \right\}_{l=2}^n 
        \atop{\scs 
        \in {\cal X}^{(n-1)m}
        }
    }   
\prod_{l=2}^n P_{A_l^m}(a_l^m)      
=|\mathcal{X}|^{-2m}.
\end{align*}
Step (a) follows from (\ref{eqn:Awffc}). 
Step (b) follows from the independent 
property on $A_l^m, l=1,2,\cdots,n$ and $B^m.$
}
\newcommand{\ArgumentB}{
\begin{align}
&
\ba{l}
{\vcs} A \oplus b^{m}=
{\vct} A \oplus b^{m}=\widetilde{s}^{m}
\ea
\notag\\
&\Leftrightarrow 
\ba{l}
({\vcs}\ominus {\vct})A=0,
b^m=\ds \widetilde{s}^{m} \ominus 
\left\{\sum_{l=1}^n s_la_l^m \right\} 
\ea
\notag\\
&\Leftrightarrow 
\ba{l}
\ds a_1^m=\sum_{l=2}^n 
   \frac{t_l \ominus s_l}{s_1 \ominus t_1} a_l^m,
\ds b^m= \widetilde{s}^{m} \ominus 
\left\{ \sum_{l=1}^n s_l a_l^m \right\} 
\ea
\notag\\
&
\Leftrightarrow 
\ba{l}
\ds a_1^m=\sum_{l=2}^n 
   \frac{t_l \ominus s_l}{s_1 \ominus t_1} a_l^m,
\ds b^m= \widetilde{s}^m \oplus 
\sum_{l=2}^n
\frac{t_1s_l\ominus s_1t_l}{s_1\ominus t_1}a_l^m.
\ea
\label{eqn:Awffc}
\end{align}
Computing 
$\Pr[{\vcs} A \oplus b^{m}={\vct} A \oplus b^{m}=\widetilde{s}^{m}]$,
we have the following chain of equalities:
\begin{align*}
&\Pr[{\vcs} A \oplus b^{m}={\vct} A \oplus b^{m}=\widetilde{s}^{m}]
\\
&\MEq{a}
\Pr\left[
a_1^m=\sum_{l=2}^n\frac{t_l \ominus s_l}{s_1 \ominus t_1} a_l^m
\right.
\\
&\qquad \left.
\land
b^m=\widetilde{s}^m \oplus 
\sum_{l=2}^n
\frac{t_1s_l \ominus s_1t_l}
{s_1\ominus t_1}a_l^m
\right]
\\
&\MEq{b}
\sum_{ \scs \left\{ a_l^m \right \}_{l=2}^n 
      \atop{\scs
        \in {\cal X}^{(n-1)m}
      }
    }   
\left[\prod_{l=2}^n P_{A_l^m}(a_l^m)\right]      
P_{A_1^m}\left(\sum_{l=2}^n\frac{t_l \ominus s_l}{s_1 \ominus t_1} 
a_l^m \right)
\\
&\qquad \times 
P_{B^m}
\left(
\widetilde{s}^m \oplus
\sum_{l=2}^n
\frac{t_1s_l\ominus s_1t_l}
{s_1\ominus t_1}a_l^m
\right)
\\
&=|\mathcal{X}|^{-2m}
 \sum_{ \scs \left\{a_l^m \right\}_{l=2}^n 
        \atop{\scs 
        \in {\cal X}^{(n-1)m}
        }
    }   
\prod_{l=2}^n P_{A_l^m}(a_l^m)      
=|\mathcal{X}|^{-2m}.
\end{align*}
Step (a) follows from (\ref{eqn:Awffc}). 
Step (b) follows from the independent 
property on $A_l^m, l=1,2,\cdots,n$ and $B^m.$
}
\ArgumentB
\IEEEQED
}
We next define the decoder function 
$\psi^{(n)}: 
{\cal X}^{m} \to {\cal X}^{n}.$
To this end we define the following quantities.   
\begin{definition}{\rm
 For ${\vcxone} \in{\cal X}^{n}$,
we denote the entropy calculated
from the type $P_{{\lvcxone}}$ by 
$H({\vcxone})$. 
In other words, for a type  
$P_{\overline{X}} \in {\cal P}_n({\cal X})$ 
such that $P_{\overline{X}}=P_{{\lvcxone}}$, 
we define 
$H({\vcxone})=H(\overline{X})$.
}
\end{definition}

\noindent
\underline{\it Minimum Entropy Decoder:} \ For 
$\phi^{(n)}(x^n)=\widetilde{x}^m$, we define the 
decoder function 
$\psi^{(n)}: {\cal X}^{m}\to {\cal X}^n$ 
as follows:
$$
\psi^{(n)}(\widetilde{x}^{m})
:=\left\{\begin{array}{cl}
{\hvcxone}
   &\mbox{if } \phi^{(n)}({\hvcxone})=\widetilde{x}^{m},\\
   &\mbox{and }H({\hvcxone})
    <H( {\cvcxone})\\
   &\mbox{for all }{\cvcxone}\mbox{ such that }\\
   & \:\phi^{(n)}({\cvcxone})=\widetilde{x}^{m},\\
   & \mbox{and } 
    \:{\cvcxone} \neq {\hvcxone},
\vspace{0.2cm}\\
\mbox{arbitrary}
    & \mbox{if there is no such }{\hvcxone}
     \in{\cal X}^{n}.
\end{array}
\right.
$$

\noindent
\underline{\it Error Probability Bound:} \ In the following 
arguments we let expectations based on the random choice of 
the affine encoder $\varphi^{(n)}$ be denoted by 
${\bf E}$$[$$\cdot]$. Define 
$$
\Psi_{\overline{X}}(R):=
    {\ExP}^{-n[R-H(\overline{X})]^{+}}.
$$
Then we have the following lemma.
\begin{lemma}\label{lem:LemA}
For any $n$ and for any
$P_{\overline{X}}
\in{\cal P}_{n}({\cal X})$,
$$
{\bf E}\left[
\Xi_{\overline{X}}(\phi^{(n)},\psi^{(n)})
\right]
\leq
{\ExP}(n+1)^{|{\cal X}|}\Psi_{\overline{X}}(R).
$$
\end{lemma}

Proof of this lemma is given in Appendix \ref{apd:ProofLemA}.
\newcommand{\ProofLemA}{
\subsection{Proof of Lemma \ref{lem:LemA}}
\label{apd:ProofLemA}

\begin{IEEEproof}[Proof of Lemma \ref{lem:LemA}] 
For ${{\vcxone}}\in {\cal X}^{n}$ we set 
\begin{align*}
B({\vcxone}  )&=
\Bigl\{\,({\cvcxone}):\,
H({\cvcxone})
\leq H({{\vcxone}})\,,\:
P_{{\clvcxone}}=P_{{\lvcxone}}\,
\Bigr\},
\end{align*}
Using parts a) and b) of Lemma \ref{lem:Lem1}, we have 
following inequalities:

\begin{align}
|B({\vcxone})|
&\leq (n+1)^{|{\cal X}|}{\ExP}^{nH({\lvcxone})},
\label{eqn:b1}
\end{align}
On an upper bound of ${\bf E}[\Xi_{{\lvcxone}}
(\phi^{(n)},\psi^{(n)})]$,
we have the following chain of inequalities:
\begin{align*}
&  {\bf E}[\Xi_{{{\lvcxone}}}(\phi^{(n)},\psi^{(n)})]
\leq\sum_{\scs
      {\clvcxone} \in B({{\lvcxone}}),
       \atop{\scs 
             {\clvcxone} \neq {\lvcxone} 
            }
     }
\Pr\bigl\{\phi^{(n)}({\cvcxone})
          =\phi^{(n)}({\vcxone})\bigr\}
\\
&\MLeq{a}\sum_{ {\clvcxone} \in 
      B({\lvcxone})}\frac{1}{|{\cal X}|^{m}}
= \frac{|B({\vcxone}  )|}{|{\cal X}|^{m}}
\MLeq{b} {\ExP}(n+1)^{|{\cal X}|}{\ExP}^{-n[R-H({{\lvcxone}} )]}.
\end{align*}
Step (a) follows from Lemma \ref{lem:good_set} part a). 
Step (b) follows from (\ref{eqn:b1}) 
and $|{\cal X}|^{m} \geq {\ExP}^{nR-1}$. 
On the other hand we have the obvious bound 
${\bf E}[\Xi_{{{\lvcxone}}}(\phi^{(n)},\psi^{(n)})]\leq 1$.
Hence we have 
\begin{align*}
&{\bf E}[\Xi_{{{\lvcxone}}}(\phi^{(n)},\psi^{(n)})]
\\
& \leq {\ExP}(n+1)^{|{\cal X}|}
\left\{{\ExP}^{-n[R-H({{\lvcxone}})]^{+}}\right\}.
\end{align*}
Hence we have
\begin{align*}
&{\bf E}[\Xi_{\overline{X}_1\overline{X}_2}(\phi^{(n)},\psi^{(n)})]
={\bf E}
\left[
\frac{1}{|T^n_{\overline{X}}|}
\sum_{{\lvcxone}
\in T^n_{\overline{X}}}
\Xi_{{{\lvcxone}}}(\phi^{(n)},\psi^{(n)})
\right]
\\
&= \frac{1}{|T^n_{\overline{X}}|}
  \sum_{{\lvcxone} \in T^n_{\overline{X} }}
 {\bf E}[\Xi_{{\lvcxone}}(\phi^{(n)},\psi^{(n)})]
\\
&\leq {\ExP}(n+1)^{|{\cal X}|}
\left\{
 {\ExP}^{-n[R-H(\overline{X})]^{+}}
\right\},
\end{align*}
completing the proof.
\end{IEEEproof}
}

\noindent
\underline{\it Estimation of Approximation Error:} \ 
Define 
\begin{align*}
&\Theta(R,\varphi_{\cal A}^{(n)}|p_{K^n},W^n) 
:=\sum_{(a, k^n) \in {\cal M}_{\cal A}^{(n)} \times {\cal X}^n} 
{p_{M_{\cal A}^{(n)} K^n}(a,k^n)}
\\
&\quad \times \log\left[ 1+ ({\rm e}^{nR}-1) 
p_{K^n|M_{\cal A}^{(n)}}(k^n|a)\right].
\end{align*}

Then we have the following lemma. 
\begin{lemma}\label{lem:LemB} \ \ For any $n,m$ satisfying 
$(m/n) \log |{\cal X}|$ $\leq R$, we have
\begin{align}
& \E \left[ D \!\!\left. \left. \left. 
\left(p_{\tilde{K}^m|M_{{\cal A}}^{(n)}}
\right|\right|
p_{V^m} \right|p_{M_{\cal A}^{(n)}}\right)\right]
\notag\\
& \leq \Theta(R,\varphi_{\cal A}^{(n)}|p_{K^n},W^n).
\label{eqn:Lem2aS}
\end{align}
\end{lemma}

Proof of this lemma is given in Appendix \ref{apd:ProofLemB}. 
From the bound (\ref{eqn:Lem2aS}) 
in Lemma (\ref{lem:LemB}), we know that  
the quantity $\Theta(R,\varphi_{\cal A}^{(n)}|p_{K^n},W^n)$
serves as an upper bound of 
the ensemble average of the conditional divergence 
$D(p_{\tilde{K}^m|M_{{\cal A}}^{(n)}}$ $
||p_{V^m}| p_{M_{\cal A}^{(n)}}).$
Hayashi \cite{hayashi:10} obtained the same upper bound 
of the ensemble average of the conditional divergence 
for an ensemble of universal${}_2$ 
functions. In this paper we prove the bound (\ref{eqn:Lem2aS}) 
for an ensemble of affine encoders. To derive this bound 
we need to use Lemma \ref{lem:good_set} parts b) and c), 
the two important properties which a class of random affine 
encoders satisfies.
\newcommand{\ProofLemB}{
\subsection{Proof of Lemma \ref{lem:LemB}}
\label{apd:ProofLemB}


In this appendix we prove Lemma \ref{lem:LemB}. This lemma 
immediately follows from the following lemma:
\begin{lemma}\label{lem:LemBb}
\ \ For any $n,m$ satisfying 
$(m/n) \log |{\cal X}|$ $\leq R$, we have  
\begin{align}
& \E \left[ D \!\!\left. \left. \left. 
\left(p_{\tilde{K}^m|M_{{\cal A}}^{(n)}}
\right|\right|
p_{V^m} \right|p_{M_{\cal A}^{(n)}}\right)\right]
\notag\\
& \leq \sum_{(a, k^n) \in {\cal M}_{\cal A}^{(n)} 
\times {\cal X}^n}{p_{M_{\cal A}^{(n)} K^n}(a,k^n)}
\notag\\
&\quad \times \log\left[ 1+ (|{\cal X}^m|-1) 
p_{K^n|M_{\cal A}^{(n)}}(k^n|a)\right]. 
\label{eqn:Lem2aT}
\end{align}
\end{lemma}

In fact, from $|{\cal X}^m|\leq {\rm e}^{nR}$ and (\ref{eqn:Lem2aT}) in 
Lemma \ref{lem:LemBb}, we have the bound (\ref{eqn:Lem2aS}) in 
Lemma \ref{lem:LemB}. Thus, we prove Lemma \ref{lem:LemBb} instead of proving 
Lemma \ref{lem:LemB}. In the following arguments, we use the following 
simplified notations: 
\begin{align*}
 k^n , K^n \in {\cal X}^n &\Longrightarrow  k , K  \in {\cal K}, 
\\
 \tilde{k}^m , \tilde{K}^m \in {\cal X}^m &\Longrightarrow l, L \in {\cal L}, 
\\
\varphi^{(n)}: {\cal X}^n \to {\cal X}^m &\Longrightarrow 
\varphi: {\cal K} \to {\cal L}, 
\\
\varphi^{(n)}(k^n)=k^n A +b^m 
& \Longrightarrow  \varphi(k) =k A+b, 
\\
 V^m \in{\cal X}^m & \Longrightarrow  V \in {\cal L},
\\ 
 M_{{\cal A}}^{(n)} \in {\cal M}_{{\cal A}}^{(n)}
& \Longrightarrow  M \in {\cal M}.
\end{align*}
We define
$$
\chi_{\varphi(k),l}=
\left\{
\ba{l}
1,\mbox{ if }\varphi(k)=l, 
\\
0,\mbox{ if }\varphi(k) \neq l. 
\ea
\right.
$$
Then, the conditional distribution
of the random variable $L= L_{\varphi}$ 
for given $M=a \in {\cal M}$
is 
\begin{align*}
&p_{L|M}(l|a)
=\sum_{k \in {\cal K}} p_{K|M} (k|a)
\chi_{\varphi(k),l} \mbox{ for }l \in {\cal L}.
\end{align*}
Define 
\begin{align*}
& \Upsilon_{ \varphi(k),l}
\defeq \chi_{\varphi(k),l}
\log \hugebl |{\cal L}|\hugel
\sum_{k^{\prime} \in {\cal K}} p_{K|M}(k^{\prime}|a)
\chi_{\varphi(k^{\prime}),l}
\huger \hugebr.
\end{align*}
Then the conditional divergence  between 
$p_{L|M}$ and $p_{V}$ for given $M$ is given by 
\begin{align}
&\left. \left. \left. D \left(p_{L|M}\right|\right|
p_{V} \right | p_M \right)
=\sum_{(a,k)\in {\cal M}\times {\cal K}} 
 \sum_{l\in {\cal L}}p_{MK} (a,k) 
\Upsilon_{\varphi(k),l}.
\end{align}
The quantity 
$\Upsilon_{\varphi(k),l} $ has the following form:
\begin{align}
& \Upsilon_{\varphi(k),l}=\chi_{\varphi(k),l}
\log \Biggl\{ |{\cal L}|\Biggl(p_{K |M}(k|a)\chi_{\varphi(k),l}
\notag\\
& \left. \left. \qquad + 
\sum_{k^{\prime} \in \{k\}^{\rm c}} 
p_{K |M}(k^{\prime} |a) 
\chi_{\varphi(k^{\prime}),l}
\right)\right\}.
\label{eqn:AzzxW}
\end{align}
The above form is useful for computing 
$\E[ \Upsilon_{\varphi(k),l}]$. 

\begin{IEEEproof}[Proof of Lemma \ref{lem:LemBb}] 
Taking expectation of both  side of (\ref{eqn:AzzxW})
with respect to the random choice of the entry of the matrix $A$ and 
the vector $b$ representing the affine encoder $\varphi$, 
we have 
\begin{align}
&\left. \left. \left. \E \left[ D \left(p_{L|M }
\right|\right|
p_{V} \right| p_{M}\right)\right]
\notag\\
&=\sum_{(a,k) \in {\cal M}\times {\cal K}}
 \sum_{l \in {\cal L}}p_{MK}(a,k)
\E \left[\Upsilon_{\varphi(k),l}\right].
\label{eqn:Zdxxp}
\end{align}
To compute the expectation 
$\E \left [\Upsilon_{\varphi(k),l}\right]$, 
we introduce an expectation operator useful for the computation.
Let $\E_{\varphi(k)=l_k}[\cdot]$ 
be an expectation operator based on the conditional probability measures
${\rm Pr}(\cdot|\varphi(k)=l_k)$.
Using this expectation operator, the quantity 
$\E \left[\Upsilon_{\varphi(k),l}\right]$ 
can be written as 
\begin{align}
&\E \left[\Upsilon_{\varphi(k),l}\right]
=\sum_{ l_k \in {\cal L}}{\rm Pr} \left(\varphi(k)=l_k\right)
\E_{\varphi(k)=l_k}\left[\Upsilon_{l_k,l}\right].
\label{eqn:SddXPP}
\end{align}
Note that 
\beq
\Upsilon_{l_k,l}
=\left\{
\ba{l}
1, \mbox{ if } l_k=l,\\
0, \mbox{ otherwise.}
\ea
\right.
\label{eqn:SdXl}
\eeq
From (\ref{eqn:SddXPP}) and (\ref{eqn:SdXl}), we have 
\begin{align}
&\E \left[\Upsilon_{\varphi(k),l}\right]
={\rm Pr} \left(\varphi(k)=l\right)\E_{\varphi(k)=l}\left[\Upsilon_{l,l}\right]
\notag\\
&=\frac{1}{|{\cal L}|}\E_{\varphi(k)=l}\left[\Upsilon_{l,l}\right].
\label{eqn:ASdff}
\end{align}
Using (\ref{eqn:AzzxW}), the expectation 
$\E_{\varphi(k)=l}\left[\Upsilon_{l,l}\right]$ 
can be written as 
\begin{align}
& \E_{\varphi(k)=l} \left[ \Upsilon_{l,l} \right] 
=\E_{\varphi(k)=l} 
  \Biggl[
  \log \Biggl\{ |{\cal L}|\Biggl(p_{K |M}(k|a)
  \notag\\
& \left.\left.\left. 
\qquad + 
\sum_{k^{\prime} \in \{k\}^{\rm c}} 
p_{K |M}(k^{\prime} |a) 
\chi_{\varphi(k^{\prime}),l}
\right)\right\}\right].
\label{eqn:AzzxWcc}
\end{align}
Applying Jensen's inequality to the right member of (\ref{eqn:AzzxWcc}), 
we obtain the following upper bound of 
$\E_{\varphi(k)=l} \left[\Upsilon_{l,l}\right]$:
\begin{align}
& \E_{\varphi(k)=l} \left[\Upsilon_{l,l}\right]
\leq 
\log \Biggl\{ |{\cal L}|\Biggl(p_{K |M}(k|a)
\notag\\
& \left.\left.
\qquad + \sum_{k^{\prime} \in \{k\}^{\rm c}} 
        p_{K |M}(k^{\prime} |a)
\E_{\varphi(k)=l}\left[\chi_{\varphi(k^{\prime}),l}\right] 
\right)\right\}
\notag\\
&\MEq{a} 
\log \Biggl\{|{\cal L}|\Biggl(
        p_{K |M}(k|a) + \sum_{k^{\prime} \in \{k\}^{\rm c}} 
        p_{K |M}(k^{\prime} |a)\frac{1}{|{\cal L}|}
\Biggr)\Biggr\}
\notag\\
&= \log \left\{1+ (|{\cal L}|-1)p_{K |M}(k|a)\right\}.
\label{eqn:AzzxWfd}
\end{align}
Step (a) follows from that 
by Lemma \ref{lem:good_set} parts b) and  c), 
\begin{align*}
\E_{ \varphi(k)=l}\left[\chi_{\varphi(k^{\prime}),l}\right] 
 &={\rm Pr} (\varphi(k^{\prime})=l|\varphi(k)=l)=\frac{1}{|{\cal L}|}.
\end{align*}
From (\ref{eqn:Zdxxp}), (\ref{eqn:ASdff}), and (\ref{eqn:AzzxWfd}), 
we have the bound  (\ref{eqn:Lem2aT}) 
in Lemma \ref{lem:LemBb}.
\end{IEEEproof}
}
From Lemmas \ref{lem:MIandDiv} and \ref{lem:LemB}, we 
have the following corollary. 
\begin{corollary}\label{cor:Szz}
\begin{align*}
{\bf E}\left[
\Delta_{n}(\varphi^{(n)},\varphi_{\cal A}^{(n)}|
p^n_{X}, p^n_{K},W^n)\right]
\leq  \Theta(R,\varphi_{\cal A}^{(n)}|p^n_{K},W^n).
\end{align*}
\end{corollary}

\noindent
\underline{\it Existence of 
Good Universal Code $(\varphi^{(n)},\psi^{(n)})$:} 

From Lemma \ref{lem:LemA} and Corollary \ref{cor:Szz}, 
we have the following lemma stating an existence 
of good universal code $(\varphi^{(n)},\psi^{(n)})$. 

\begin{lemma}\label{lem:UnivCodeBounda} There exists at 
least one deterministic code  
$(\varphi^{(n)},\psi^{(n)})$ satisfying 
$(m/n)\log |{\cal X}|\leq R$, such that 
for any 
$p_{\overline{X}}$ $\in {\cal P}_n({\cal X})$,
\begin{align*}
&\Xi_{\overline{X}}(\phi^{(n)},\psi^{(n)})
\\
&\leq {\ExP}(n+1)^{|{\cal X}|}\{(n+1)^{|{\cal X}|}+1\}
\Psi_{\overline{X}}(R).
\end{align*}
Furthermore, for any $\varphi_{\cal A}^{(n)} 
\in {\cal F}_{\cal A}^{(n)}(R_{\cal A})$, we have  
\begin{align*}
&\Delta_{n}(\varphi^{(n)},\varphi_{\cal A}^{(n)}|
p^n_{X}, p^n_{K},W^n)
\\
& \leq \{(n+1)^{|{\cal X}|}+1\}
\Theta(R,\varphi_{\cal A}^{(n)}|p^n_{K},W^n).
\end{align*}
\end{lemma}
\begin{IEEEproof}
We have the following chain of inequalities:
\begin{align}
&{\bf E}
\left[
\sum_{
p_{\overline{X}} 
\in {\cal P}_n({\cal X})
}
\frac{\Xi_{\overline{X}}(\phi^{(n)},\psi^{(n)})}
{{\ExP}(n+1)^{|{\cal X}|}\Psi_{\overline{X}}(R)}
\right.
\notag\\
&\qquad 
+\frac{\Delta_n(\varphi^{(n)},\varphi_{\cal A}^{(n)}|
 p_{X}^n,p_{K}^n,W^n)}
{\Theta(R,\varphi_{\cal A}^{(n)}|p^n_{K},W^n)}
\Hugebr
\notag\\
&=
\sum_{
p_{\overline{X}} 
\in {\cal P}_n({\cal X})
}
\frac{{\bf E}\left[
\Xi_{\overline{X}}(\phi^{(n)},\psi^{(n)})
\right]}
{{\ExP}(n+1)^{|{\cal X}|}\Psi_{\overline{X}}(R)}
\notag\\
&\quad +
\frac{{\bf E}\left[\Delta_n(\varphi^{(n)},
\varphi_{\cal A}^{(n)}|p_{X}^n,p_{K}^n,W^n)\right]}
{\Theta(R,\varphi_{\cal A}^{(n)}|p^n_{K},W^n)}
\notag\\
&\MLeq{a} 
\sum_{p_{\overline{X}}\in {\cal P}_n({\cal X})}1
+1=|{\cal P}_n({\cal X})|+1
\MLeq{b} (n+1)^{|{\cal X}|}+1.
\notag
\end{align}
Step (a) follows from Lemma \ref{lem:LemA} and 
Corollary \ref{cor:Szz}. Step (b) follows 
from Lemma \ref{lem:Lem1} part a).
Hence there exists at least one deterministic code
$(\varphi^{(n)},\psi^{(n)})$ such that
\begin{align}
&\sum_{
p_{\overline{X}} 
\in {\cal P}_n({\cal X})
}
\frac{\Xi_{\overline{X}}(\phi^{(n)},\psi^{(n)})}
{{\ExP}(n+1)^{|{\cal X}|}\Psi_{\overline{X}}(R)}
\notag\\
&\quad 
+\frac{\Delta_n(\varphi^{(n)},\varphi_{\cal A}^{(n)}
|p_{X}^n,p_{K}^n,W^n)}
{\Theta(R,\varphi_{\cal A}^{(n)}|p^n_{K},W^n)}
\leq (n+1)^{|{\cal X}|}+1,
\notag
\end{align}
from which we have that 
\begin{align*}
&\frac{\Xi_{\overline{X}}(\phi^{(n)},\psi^{(n)})}
{{\ExP}(n+1)^{|{\cal X}|}\Psi_{\overline{X}}(R)}
\leq (n+1)^{|{\cal X}|}+1,
\end{align*}
for any $p_{\overline{X}}\in {\cal P}_n({\cal X})$.
Furthermore, we have that for any 
$\varphi_{\cal A}^{(n)}
\in {\cal F}_{\cal A}^{(n)}(R_{\cal A})$,
\begin{align*}
&\frac{\Delta_n(\varphi^{(n)},\varphi_{\cal A}^{(n)}
|p_{X}^n,p_{K}^n,W^n)}
{{\Theta(R,\varphi_{\cal A}^{(n)}|p^n_{K},W^n)}}
\leq (n+1)^{|{\cal X}|}+1,
\end{align*}
completing the proof.
\end{IEEEproof}

%

\begin{proposition}\label{pro:UnivCodeBound} 
For any $R_{\cal A}, R>0$, and 
any $(p_K,W)$, there exists a sequence of mappings 
$\{(\varphi^{(n)}, \psi^{(n)}) \}_{n=1}^{\infty}$
such that for any $p_X \in {\cal P}({\cal X})$, 
we have 
\begin{align}
& \frac {1}{n} 
\log |{\cal X}^{m}|= \frac {m}{n} \log |{\cal X}|\leq R,
\notag\\
& p_{\rm e}(\phi^{(n)},\psi^{(n)}|p_{X}^n) \leq 
{\ExP}(n+1)^{2|{\cal X}|}\{(n+1)^{|{\cal X}|}+1\}  
\notag\\
&\quad \quad \qquad \qquad \qquad 
\times {\ExP}^{-n[E(R|p_{X})]} \label{eqn:mainThErrBa}
\end{align}
and for any eavesdropper $\A$ with $\varphi_{\A}$ satisfying
$\varphi_{\A}^{(n)} \in {\cal F}_{\A}^{(n)}(R_{\A})$, we have
\begin{align}
& \Delta^{(n)}(\varphi^{(n)},\varphi_{\A}^{(n)}|p_{X}^n,p_{K}^n,W^n)
\notag\\
&\leq \{(n+1)^{|{\cal X}|}+1\}
\Theta(R,\varphi_{\cal A}^{(n)}|p^n_{K},W^n).
\label{eqn:mainThSecBa}
\end{align}
\end{proposition}

\begin{IEEEproof}
By Lemma \ref{lem:UnivCodeBounda}, there exists 
$(\varphi^{(n)},$ $\psi^{(n)})$ satisfying 
$(m/n)\log |{\cal X}|\leq R$, such that 
for any $p_{\overline{X}}$ 
$\in {\cal P}_n({\cal X})$,
\begin{align}
&\Xi_{\overline{X}}(\phi^{(n)},\psi^{(n)})
\notag\\
&
\leq {\ExP}(n+1)^{|{\cal X}|}\{(n+1)^{|{\cal X}|}+1\}
\Psi_{\overline{X}}(R).
\label{eqn:aaSSS}
\end{align}
Furthermore for any $\varphi_{\cal A}^{(n)}
\in {\cal F}_{\cal A}^{(n)}(R_{\cal A})$,
\begin{align}
&\Delta_n(\varphi^{(n)},\varphi_{\cal A}^{(n)}
|p_{X}^n,p_{K}^n,W^n)
\notag\\
&\leq \{(n+1)^{|{\cal X}|}+1\}
\Theta(R,\varphi_{\cal A}^{(n)}|p^n_{K},W^n).
\label{eqn:abSSSz}
\end{align}
The bound (\ref{eqn:mainThSecBa}) in 
Proposition \ref{pro:UnivCodeBound}
has already been proved in (\ref{eqn:abSSSz}). 
Hence it suffices to prove the bound (\ref{eqn:mainThErrBa}) 
in Proposition \ref{pro:UnivCodeBound} to complete the proof.
On an upper bound of $p_{\rm e}(\phi^{(n)},\psi^{(n)}|p_{X}^n)$,  
we have the following chain of inequalities:  
\begin{align*}
& p_{\rm e}(\phi^{(n)},\psi^{(n)}|p_{X}^n)  
\\
&\MLeq{a} {\ExP}(n+1)^{|{\cal X}|} \{ (n+1)^{|{\cal X}|}+1\}
\\
&\quad \times \sum_{p_{\overline{X}}
   \in {\cal P}_n({\cal X}) }
   \Psi_{\overline{X}}(R)
   {\ExP}^{-nD(p_{\overline{X}}||p_{X})}
\\
&\leq {\ExP}(n+1)^{|{\cal X}|}\{ (n+1)^{|{\cal X}|}+1\}
|{\cal P}_n({\cal X})|{\ExP}^{-n[E(R|p_{X})]} 
\\
&\MLeq{c} {\ExP}(n+1)^{2|{\cal X}|}\{(n+1)^{|{\cal X}|}+1\} 
  {\ExP}^{-nE(R|p_{X})}.
\end{align*}
Step (a) follows from Lemma \ref{lem:ErBound}
and (\ref{eqn:aaSSS}).
Step (b) follows from Lemma \ref{lem:Lem1} part a).
\end{IEEEproof}


\subsection{Explicit Upper Bound of 
$\Theta(R,\varphi_{\cal A}^{(n)}|p^n_K,W^n)$} 

In this subsection we derive an explicit upper bound of 
$\Theta(R,\varphi_{\cal A}^{(n)}|p^n_{K},W^n)$ which holds
for any eavesdropper $\A$ with $\varphi_{\A}$ satisfying
$\varphi_{\A}^{(n)} \in {\cal F}_{\A}^{(n)}(R_{\A})$. 
Define 
\begin{align*}
& \wp \defeq p_{ M_{\cal A}^{(n)} Z^n K^n}\Biggl\{
  \nonumber\\
&  R \geq \left.
 \frac{1}{n}\log\frac{1}{p_{K^n|M_{\cal A}^{(n)}}(K^n|M_{\cal A}^{(n)})}
 -\eta
 \right\}
\end{align*}
Then we have the following lemma.
\begin{lemma}\label{lem:ThetaBound}
For any $\eta>0$ and for any eavesdropper $\A$ with 
$\varphi_{\A}$ satisfying
$\varphi_{\A}^{(n)} \in {\cal F}_{\A}^{(n)}(R_{\A})$, we have 
\beqa
& &\Theta(R,\varphi_{\cal A}^{(n)}|p^n_K,W^n)
\leq 
nR \cdot \wp +{\rm e}^{-n\eta}.
\label{eqn:azca}
\eeqa 
Specificall if $n \geq 1/R$, we have
\beqa
& &(nR)^{-1}\Theta(R,\varphi_{\cal A}^{(n)}|p^n_K,W^n)
\leq \wp +{\rm e}^{-n\eta}.
\label{eqn:Zddaa} 
\eeqa
\end{lemma}

\begin{IEEEproof}
We first observe that
\begin{align}
&\Theta(R,\varphi_{\cal A}^{(n)}|p^n_K,W^n)
\notag\\
&={\rm E}\left[
\log \left\{ 1+ ({\rm e}^{nR}-1) p_{K^n|M_{\cal A}^{(n)}}
(K^n|M_{\cal A}^{(n)})\right\}\right].
\label{eqn:AdttA}
\end{align}
We further observe the following: 
\begin{align}
&R < \frac{1}{n} \log 
\frac{1}{p_{K^n|M_{\cal A}^{(n)}}(K^n|M_{\cal A}^{(n)})}-\eta
\notag\\
&\Leftrightarrow
{\rm e}^{nR} p_{K^n|M_{\cal A}^{(n)}}(K^n|M_{\cal A}^{(n)})
< {\rm e}^{-n\eta}
\notag \\
&\Rightarrow
\log \left\{ 1+ {\rm e}^{nR} p_{K^n|M_{\cal A}^{(n)}}
(K^n|M_{\cal A}^{(n)})\right\}
\notag \\
&\quad \leq \log\left(1+{\rm e}^{-n\eta}\right)
\notag \\
&\MRarrow{a}
\log \left\{ 1+ {\rm e}^{nR} p_{K^n|M_{\cal A}^{(n)}}
(K^n|M_{\cal A}^{(n)})\right\} \leq {\rm e}^{-n\eta}
\notag \\
&\Rightarrow
\log \left\{ 1+ ({\rm e}^{nR}-1)p_{K^n|M_{\cal A}^{(n)}}
(K^n|M_{\cal A}^{(n)})\right\}
\notag\\
&\quad \leq {\rm e}^{-n\eta}.
\label{eqn:AdttB}
\end{align}
Step (a) follows from $\log (1+a)\leq a$.
We also note that
\begin{align}
&\log \left\{ 
1+ ({\rm e}^{nR}-1)
p_{K^n| M_{\cal A}^{(n)}}(K^n|M_{\cal A}^{(n)})
     \right\}
\notag\\
& 
\leq \log [{\rm e}^{nR}]=nR.
\label{eqn:AdttC}
\end{align}
From (\ref{eqn:AdttA}),
     (\ref{eqn:AdttB}),
     (\ref{eqn:AdttC}), we have
the bound (\ref{eqn:azca}) in Lemma \ref{lem:ThetaBound}.
\end{IEEEproof}

On upper bound of $\wp$, we have the following lemma.
\begin{lemma}
\label{lem:Ohzzz}
For any $\eta>0$ and for any eavesdropper 
$\A$ with $\varphi_{\A}$ satisfying
$\varphi_{\A}^{(n)} \in {\cal F}_{\A}^{(n)}(R_{\A})$,
we have $\wp\leq \tilde{\wp}$, where

\begin{align}
&\!\!\!\!\!\!\!\!
\tilde{\wp}\defeq
p_{M_{\cal A}^{(n)}Z^nK^n}\Biggl\{
\notag\\
0& \geq \frac{1}{n}\log
\frac{\hat{q}_{M_{\cal A}^{(n)}Z^nK^n}(M_{\cal A}^{(n)},Z^n,K^n)}
{p_{M_{\cal A}^{(n)}Z^nK^n}(M_{\cal A}^{(n)},Z^n,K^n)}-\eta,
\label{eqn:asppa}\\
0& \geq \frac{1}{n}\log \frac{ q_{{Z}^n}(Z^n)}{p_{Z^n}(Z^n)}-\eta,
\label{eqn:asppb}\\
R_{\cal A}&\geq \ds \frac{1}{n}\log
\frac{p_{Z^n|M_{\cal A}^{(n)}}(Z^n|M_{\cal A}^{(n)})}
{p_{Z^n}(Z^n)}-\eta,
\notag\\
R&\geq \ds \frac{1}{n}\log 
\frac{1}{p_{K^n|M_{\cal A}^{(n)}}(K^n|M_{\cal A}^{(n)})}
-\eta
\Biggr\}
+3{\rm e}^{-n\eta}. 
\label{eqn:azsad}
\end{align}
The probability distributions appearing in the two inequalities 
(\ref{eqn:asppa}) and (\ref{eqn:asppb}) in the right members 
of (\ref{eqn:azsad}) have a property that 
we can select them arbitrary. 
In (\ref{eqn:asppa}), we can choose any probability 
distribution $\hat{q}_{M_{\cal A}^{(n)}Z^nK^n}$ on 
${\cal M}_{\cal A}^{(n)}$$\times{\cal Z}^n$$\times{\cal X}^n$. 
In (\ref{eqn:asppb}), we can choose any 
distribution $q_{Z^n}$ on ${\cal Z}^n$. 
\end{lemma}

Proof of this lemma is given in Appendix \ref{sub:Apda}.
\newcommand{\Apda}{
\subsection{
Proof of Lemma \ref{lem:Ohzzz}
}\label{sub:Apda}

To prove Lemma \ref{lem:Ohzzz}, we prepare a lemma.
For simplicity of notation, set 
$|{\cal M}_{\cal A}^{(n)}|=M_{\cal A}$.
Define
$$
{\cal B}_n
\defeq 
\left \{(a,z^{n},k^{n}): 
\frac{1}{n}\log 
\frac {p_{M_{\cal A}^{(n)}Z^nK^n}
(a,z^n,k^n)}{ \hat{q}_{M_{\cal A}^{(n)}Z^nK^n}(a,z^n,k^n)}
\geq -\eta
\right\}.
$$
Furthermore, define
\beqno
& &\tilde{\cal C}_n
\defeq 
\left\{z^n: 
\frac{1}{n}\log 
\frac {p_{Z^n}(z^n)}{q_{Z^n}(z^n)}
\geq -\eta
\right\},
\\
&&{\cal C}_n \defeq \tilde{\cal C}_n 
\times {\cal M}_{\cal A}^{(n)}\times {\cal X}^n,
  {\cal C}_n^{\rm c}
  \defeq \tilde{\cal C}_n^{\rm c}
\times {\cal M}_{\cal A}^{(n)} \times {\cal X}^n,
\\
&&\tilde{\cal D}_n\defeq
\{(a,z^n): 
\ba[t]{l}
a=\varphi_{\cal A}^{(n)}(z^n),\\ 
 p_{Z^n|M_{\cal A}^{(n)}}(z^n|a)\leq 
M_{\cal A}{\rm e}^{n\eta}p_{Z^n}(z^n)\}, 
\ea
\\
&&{\cal D}_n \defeq \tilde{\cal D}_n \times {\cal X}^n,
  {\cal D}_n^{\rm c}
  \defeq \tilde{\cal D}_n^{\rm c}\times {\cal X}^n,
\\
&&{\cal E}_n \defeq
\{(a,z^n,k^n): 
\ba[t]{l}
a=\varphi_{\cal A}^{(n)}(z^n),\\ 
p_{K^n|M_{\cal A}^{(n)}}(k^n|a) 
\geq {\rm e}^{-n(R+\eta)}\}.
\ea
\eeqno
Then we have the following lemma. 
\begin{lemma}\label{lem:zzxa}{
\beqno
& &
p_{M_{\cal A}^{(n)}Z^nK^n}
\left({\cal B}_n^{\rm c}\right)\leq {\rm e}^{-n\eta}, 
p_{M_{\cal A}^{(n)}Z^nK^n}
\left({\cal C}_n^{\rm c}\right)\leq {\rm e}^{-n\eta}, 
\\
& &p_{M_{\cal A}^{(n)}Z^nK^n}
\left({\cal D}_n^{\rm c} \right)\leq {\rm e}^{-n\eta}.
\eeqno
}
\end{lemma}

{\it Proof:} We first prove the first inequality. 
\beqno
& &p_{M_{\cal A}^{(n)}Z^nK^n}
({\cal B}_n^{\rm c})
=\sum_{(a,z^n,k^n) \in {\cal B}_n^{\rm c}}
   p_{M_{\cal A}^{(n)}Z^nK^n}(a,z^n,k^n)
\\
&\MLeq{a}&\sum_{(a,z^n,k^n)\in {\cal B}_n^{\rm c}}
{\rm e}^{-n\eta}\hat{q}_{M_{\cal A}^{(n)}Z^nK^n}(a,z^n,k^n)
\\
&=&{\rm e}^{-n\eta}q_{M_{\cal A}^{(n)}Z^nK^n}
\left({\cal B}_n^{\rm c}\right)
\leq {\rm e}^{-n\eta}.
\eeqno
Step (a) follows from the definition of ${\cal B}_n$.   
On the second inequality we have   
\beqno
& &p_{M_{\cal A}^{(n)}Z^nK^n}
({\cal C}_n^{\rm c})
= p_{Z^n}
(\tilde{\cal C}_n^{\rm c})
=\sum_{x^n\in \tilde{\cal C}_n^{\rm c} } p_{Z_n}(z^n)
\\
&\MLeq{a}&\sum_{x^n\in \tilde{\cal C}_n^{\rm c}}
{\rm e}^{-n\eta}q_{Z^n}(z^n)
={\rm e}^{-n\eta}q_{Z^n}
\left( \tilde{\cal C}_n^{\rm c}\right)
\leq {\rm e}^{-n\eta}.
\eeqno
Step (a) follows from the definition of ${\cal C}_n$.  
We finally prove the third inequality. 
\beqno
& &p_{M_{\cal A}^{(n)}Z^nK^n}
   ({\cal D}_n^{\rm c})
   =p_{M_{\cal A}^{(n)}Z^n}(\tilde{\cal D}_n^{\rm c})
\\
&=&\sum_{a \in {\cal M}_{\cal A}^{(n)}}
\sum_{\scs z^n: \varphi_{\cal A}^{(n)}(z^n)=a
\atop{\scs
      p_{Z^n}(z^n) \leq ({\rm e}^{-n\eta}/M_{\cal A})
      \atop{\scs
           \qquad \times p_{Z^n|M_{\cal A}^{(n)} }(z^n|a)
           }
     }
}
p_{Z^n}(z^n)
\\
&\leq &
\frac{{\rm e}^{-n\eta}}{ M_{\cal A} }
\sum_{a \in {\cal M}_{\cal A}^{(n)} }
\sum_{\scs z^n: \varphi_{\cal A}^{(n)}(z^n)=a
\atop{\scs
      p_{Z^n}(z^n)\leq ({\rm e}^{-n\eta}/M_{\cal A})
      \atop{\scs 
          \qquad \times p_{Z^n|M_{\cal A}^{(n)}}(z^n|a)
           }
     }
}
p_{Z^n|M_{\cal A}^{(n)}}(z^n|a)
\\
&\leq& \frac{{\rm e}^{ -n\eta }}{ M_{\cal A} } 
|{\cal M}_{\cal A}^{(n)}|
={\rm e}^{-n\eta}.
\eeqno
\hfill\IEEEQED

{\it Proof of Lemma \ref{lem:Ohzzz}:} 
By definition we have
\begin{align*}
 &p_{M_{\cal A}^{(n)}Z^nK^n}
\left(
{\cal B}_n\cap {\cal C}_n\cap {\cal D}_n
\cap {\cal E}_n
\right)
\\
=& p_{M_{\cal A}^{(n)}Z^nK^n}
\left\{
\frac{1}{n}\log 
\frac{p_{M_{\cal A}^{(n)}Z^nK^n}(M_{\cal A}^{(n)},Z^n,K^n)}
{\hat{q}_{M_{\cal A}^{(n)}Z^n K^n}(M_{\cal A}^{(n)},Z^n,K^n)}
\geq -\eta,
\right.
\\
& \qquad \quad \quad\:0 \geq \frac{1}{n}\log \frac{ q_{Z^n}(Z^n)}{p_{Z^n}(Z^n)}-\eta,
\\
& \:\:\: \frac{1}{n} \log M_{\cal A} \geq \frac{1}{n}\log
      \frac{p_{Z^n|M_{\cal A}^{(n)}}(Z^n|M_{\cal A}^{(n)})}{p_{Z^n}(Z^n)}-\eta,
\\
& \qquad \quad \quad R \geq \left.
\frac{1}{n}\log\frac{1}{p_{K^n|M_{\cal A}^{(n)}}(K^n|M_{\cal A}^{(n)})}-\eta
\right\}.
\end{align*}
Then for any $\varphi_{\cal A}^{(n)}$ 
satisfying 
$
(1/n)\log {||\varphi_{\cal A}^{(n)}||} \leq R_{\cal A},
$
we have 
\begin{align*}
 &p_{M_{\cal A}^{(n)}Z^nK^n}
\left(
{\cal B}_n\cap {\cal C}_n\cap {\cal D}_n\cap {\cal E}_n
\right)
\\
\leq & p_{M_{\cal A}^{(n)}Z^nK^n}\left\{
\frac{1}{n}\log 
\frac{ p_{M_{\cal A}^{(n)}Z^nK^n}(M_{\cal A}^{(n)},Z^n,K^n)}
{\hat{q}_{M_{\cal A}^{(n)}Z^nK^n}(M_{\cal A}^{(n)},Z^n,K^n)}
\geq -\eta, \right.
\\
& \qquad \quad 0 \geq \frac{1}{n}\log \frac{q_{Z^n}(Z^n)}{p_{Z^n}(Z^n)}-\eta,
\\
& \qquad R_{\cal A}\geq \frac{1}{n}\log
  \frac{p_{Z^n|M_{\cal A}^{(n)}}(Z^n|M_{\cal A}^{(n)})}{p_{Z^n}(Z^n)}-\eta,
\\
& \qquad R \geq \left.
\frac{1}{n}\log\frac{1}{p_{K^n|M_{\cal A}^{(n)}}(K^n|M_{\cal A}^{(n)})}-\eta
\right\}.
\end{align*}
%
%
Hence, it suffices to show 
\beqno
& &{\wp}\leq
 p_{M_{\cal A}^{(n)}Z^nK^n}\left({\cal B}_n
 \cap {\cal C}_n
 \cap {\cal D}_n \cap {\cal E}_n\right)
+3{\rm e}^{-n\eta}
\eeqno
to prove Lemma \ref{lem:Ohzzz}. 
We have the following chain of inequalities:
\begin{align*}
&{\wp}\MEq{a}p_{M_{\cal A}^{(n)}Z^nK^n}\left({\cal E}_n \right)
 \\
&=p_{M_{\cal A}^{(n)}Z^nK^n}\left(
     {\cal B}_n
\cap {\cal C}_n
\cap {\cal D}_n
\cap {\cal E}_n
\right)
\\
& \quad 
+p_{M_{\cal A}^{(n)}Z^nK^n}
\left(
\left[{\cal B}_n
 \cap {\cal C}_n
 \cap {\cal D}_n\right]^{\rm c} 
 \cap {\cal E}_n
\right)
\\
&\leq
p_{M_{\cal A}^{(n)}Z^nK^n}
\left({\cal B}_n
 \cap {\cal C}_n
 \cap {\cal D}_n
 \cap {\cal E}_n
\right)
\\
& \quad +p_{M_{\cal A}^{(n)}Z^nK^n}
   \left({\cal B}_n^{\rm c}\right)
   +p_{M_{\cal A}^{(n)}Z^nK^n}
   \left({\cal C}_n^{\rm c}\right)
\\
& \quad +p_{M_{\cal A}^{(n)}Z^nK^n}
   \left({\cal D}_n^{\rm c} \right)
\\
&\MLeq{b}
p_{M_{\cal A}^{(n)}Z^nK^n}\left(
     {\cal B}_n
\cap {\cal C}_n
\cap {\cal D}_n
\cap {\cal E}_n
\right)
+3{\rm e}^{-n\eta}=\tilde{\wp}. 
\end{align*}
Step (a) follows from the defintion of $\wp$. 
Step (b) follows from Lemma \ref{lem:zzxa}.
\hfill\IEEEQED
}

We have the following proposition. 
\begin{proposition}\label{pro:ThetaExpUpper} 
For any $\varphi_{\cal A}^{(n)} 
\in {\cal F}_{\cal A}^{(n)}(R_{\cal A})$, 
and any $n \geq 1/R$, we have 
\begin{align}
&(nR)^{-1}\Theta(R,\varphi_{\cal A}^{(n)}|p^n_K,W^n)
\leq 5{\ExP}^{-nF(R_{\cal A},R|p_K,W)}.
\label{eqn:SSssP}
\end{align}
\end{proposition}

{\it Proof:} By Lemmas \ref{lem:ThetaBound} and \ref{lem:Ohzzz}, 
we have
\begin{align}
(nR)^{-1} \Theta(R,\varphi_{\cal A}^{(n)}|p^n_K,W^n)
\leq \tilde{\wp} +{\rm e}^{-n\eta}.
\label{eqn:SSssPa}
\end{align}
The quantity $\tilde{\wp}+{\rm e}^{-n\eta}$.
is the same as the upper bound on 
the correct probability of decoding for one helper 
source coding problem in Lemma 1 in 
Oohama \cite{oohama2015exponent}(extended version). 
In a manner similar to the derivation of the exponential upper 
bound of the correct probability of decoding for one helper 
source coding problem, we can prove that 
for any $\varphi_{\cal A}^{(n)}\in {\cal F}_{\cal A}^{(n)}(R_{\cal A})$ 
and for some $\eta^*=\eta^{*}(n,R_{\cal A},R)$,
we have  
\begin{align}
&\tilde{\wp}+{\rm e}^{-n\eta^*}
\leq 5{\ExP}^{-nF(R_{\cal A},R|p_K,W)}.
\label{eqn:SSssPb} 
\end{align}
From (\ref{eqn:SSssPa}) and (\ref{eqn:SSssPb})
we have (\ref{eqn:SSssP}).
\hfill\IEEEQED

From Propositions \ref{pro:UnivCodeBound} and \ref{pro:ThetaExpUpper},
we immediately obtain Theorem \ref{Th:mainth2}.

%


\newcommand{\Omitzzz}{

\begin{corollary}
\label{cor:SzzP}
For any $\varphi_{\cal A}^{(n)} 
\in {\cal F}_{\cal A}^{(n)}(R_{\cal A})$, 
we have 
\begin{align*}
&{\bf E}\left[
\Delta_{n}(\varphi^{(n)},\varphi_{\cal A}^{(n)}|
p^n_{X}, p^n_{K},W^n)\right]
\\
& \leq  5nR{\ExP}^{-nF(R_{\cal A},R|p_K,W)}. 
\end{align*}
\end{corollary}

\begin{IEEEproof}[Proof of Theorem \ref{Th:mainth2}]
By Lemma \ref{lem:UnivCodeBound}, there exists 
$(\varphi^{(n)},$ $\psi^{(n)})$ satisfying 
$(m/n)\log |{\cal X}|\leq R$, such that 
for any $p_{\overline{X}}$ 
$\in {\cal P}_n({\cal X})$,
\begin{align}
&\Xi_{\overline{X}}(\phi^{(n)},\psi^{(n)})
\notag\\
&
\leq {\ExP}(n+1)^{|{\cal X}|}\{(n+1)^{|{\cal X}|}+1\}
\Psi_{\overline{X}}(R).
\label{eqn:aaSSS}
\end{align}
Furthermore for any $\varphi_{\cal A}^{(n)}
\in {\cal F}_{\cal A}^{(n)}(R_{\cal A})$,
\begin{align}
&\Delta_n(\varphi^{(n)},\varphi_{\cal A}^{(n)}
|p_{X}^n,p_{K}^n,W^n)
\notag\\
&\leq 5nR\{(n+1)^{|{\cal X}|}+1\}{\ExP}^{-nF(R_{\cal A},R|p_K,W)}.
\label{eqn:abSSS}
\end{align}
The bound (\ref{eqn:mainThSecB}) in Theorem \ref{Th:mainth2}
has already been proved in (\ref{eqn:abSSS}). 
Hence it suffices to prove the bound (\ref{eqn:mainThErrB}) 
in Theorem \ref{Th:mainth2} to complete the proof.
On an upper bound of $p_{\rm e}(\phi^{(n)},\psi^{(n)}|p_{X}^n)$,  
we have the following chain of inequalities:  

\begin{align*}
& p_{\rm e}(\phi^{(n)},\psi^{(n)}|p_{X}^n)  
\\
&\MLeq{a} {\ExP}(n+1)^{|{\cal X}|} \{ (n+1)^{|{\cal X}|}+1\}
\\
&\quad \times \sum_{p_{\overline{X}}
   \in {\cal P}_n({\cal X}) }
   \Psi_{\overline{X}}(R)
   {\ExP}^{-nD(p_{\overline{X}}||p_{X})}
\\
&\leq {\ExP}(n+1)^{|{\cal X}|}\{ (n+1)^{|{\cal X}|}+1\}
|{\cal P}_n({\cal X})|{\ExP}^{-n[E(R|p_{X})]} 
\\
&\MLeq{c} {\ExP}(n+1)^{2|{\cal X}|}\{(n+1)^{|{\cal X}|}+1\} 
  {\ExP}^{-nE(R|p_{X})}
\\
&={\ExP}^{-n[E(R|p_{X})-\delta_{1,n}]}.
\end{align*}
Step (a) follows from Lemma \ref{lem:ErBound}
and (\ref{eqn:aaSSS}).
Step (b) follows from Lemma \ref{lem:Lem1} part a).
\end{IEEEproof}
}

\appendix

\ProofLemAA
\ProofLemA
\ProofLemB
\Apda

\bibliographystyle{IEEEtran}

\bibliography{RefEdByOh}

\end{document}